\documentclass[acmsmall]{acmart}
\usepackage[normalem]{ulem}

\setcopyright{acmlicensed}
\acmJournal{PACMHCI}
\acmYear{2022} \acmVolume{6} \acmNumber{CSCW2} \acmArticle{370} \acmMonth{11} \acmPrice{15.00}\acmDOI{10.1145/3555095}

\received{July 2021}
\received[revised]{November 2021}
\received[accepted]{April 2022}

\newcommand{\cut}[1]{}
\newcommand{\rredit}[1]{#1}
\newcommand{\rrdelete}[1]{}
\newcommand{\rreditsec}[1]{#1}
\newcommand{\rrdeletesec}[1]{}
\newcommand{\rrdeletesubsec}[1]{}
\newif\ifmarkup
\markupfalse
\definecolor{burgundy}{RGB}{138, 14, 51}
\newcommand{\mredit}[1]{{\textcolor{purple}{#1}}}
\newcommand{\mrdelete}[1]{{\textcolor{gray}{\sout{#1}}}}

\ifmarkup
\else
\renewcommand{\mredit}[1]{#1}
\renewcommand{\mrdelete}[1]{}
\fi

\newif\ifshowcomments
\showcommentsfalse
\newcommand\todo{\textcolor{red}}
\newcommand{\jpc}[1]{{\textcolor{teal}{JPC:#1}}}
\newcommand{\cd}[1]{\textcolor{blue}{CD:#1}}
\newcommand{\kl}[1]{\textcolor{orange}{KL:#1}}

\definecolor{jspurple}{rgb}{147, 0, 191}
\newcommand{\js}[1]{{\textcolor{jspurple}{#1}}}

\ifshowcomments
\else
\renewcommand\todo[1]{}
\renewcommand{\jpc}[1]{}
\renewcommand{\cd}[1]{}
\renewcommand{\js}[1]{}
\renewcommand{\kl}[1]{}
\fi

\begin{document}

\title[Proactive Moderation of Online Discussions]{Proactive Moderation of Online Discussions: Existing Practices and the Potential for Algorithmic Support}

\author{Charlotte Schluger}
\email{jes543@cornell.edu}
\affiliation{%
  \institution{Cornell University}
  \city{Ithaca}
  \state{NY}
  \country{USA}
  \postcode{14850}
}
\author{Jonathan P. Chang}
\email{jpc362@cornell.edu}
\affiliation{%
  \institution{Cornell University}
  \city{Ithaca}
  \state{NY}
  \country{USA}
  \postcode{14850}
}
\author{Cristian Danescu-Niculescu-Mizil}
\email{cristian@cs.cornell.edu}
\affiliation{%
  \institution{Cornell University}
  \city{Ithaca}
  \state{NY}
  \country{USA}
  \postcode{14850}
}
\author{Karen Levy}
\email{karen.levy@cornell.edu}
\affiliation{%
  \institution{Cornell University}
  \city{Ithaca}
  \state{NY}
  \country{USA}
  \postcode{14850}
}

\renewcommand{\shortauthors}{Charlotte Schluger et al.}

\begin{abstract}

To address the widespread problem of uncivil behavior, many online discussion platforms employ human moderators to take action against objectionable content, such as removing it or placing sanctions on its authors. This \textit{reactive} paradigm of taking action against already-posted antisocial content is currently the most common form of moderation, and has accordingly underpinned many recent efforts at introducing automation into the moderation process. Comparatively less work has been done to understand other moderation paradigms---such as \textit{proactively} discouraging the emergence of antisocial behavior rather than reacting to it---and the role algorithmic support can play in these paradigms. \rredit{In this work, we investigate such a proactive framework for moderation in a case study of a collaborative setting: Wikipedia Talk Pages.  We employ a mixed methods approach, combining qualitative and design components for a holistic analysis.} Through interviews with moderators, we find that despite a lack of technical and social support, moderators already engage in a number of proactive moderation behaviors, such as preemptively intervening in conversations to keep them on track. Further, we explore how automation could assist with this existing proactive moderation workflow by building a prototype tool, presenting it to moderators, and examining how the assistance it provides  might fit  into their workflow. The resulting feedback uncovers both strengths and drawbacks of the prototype tool and suggests concrete steps towards further developing such assisting technology so it can most effectively support moderators in their existing proactive moderation workflow.
\end{abstract}

\ccsdesc[300]{Human-centered computing~Interactive systems and tools}
\ccsdesc[500]{Human-centered computing~Collaborative and social computing systems and tools}
\ccsdesc[500]{Computing methodologies~Natural language processing}

\keywords{antisocial behavior, content moderation, algorithmic assistance, hybrid systems}

\maketitle

\section{Introduction}
\label{sec:intro}
\rredit{Online discussion platforms enable new forms of interaction and knowledge sharing and have become key in supporting online collaboration.}
\rredit{However, incivility challenges their benefits,} harming the mental and emotional health of individuals who are exposed to antisocial behavior or targeted by personal attacks \cite{raskauskas_involvement_2007,akbulut_cyberbullying_2010}, 
\rredit{reducing engagement \cite{collier_conflict_2012}, or distracting from the underlying goals of the discussion \cite{arazy_stay_2013}}.
As a result most online platforms use moderation to keep discussions within their community guidelines.

Moderation of public discussion platforms traditionally consists of a process in which human moderators respond to instances of antisocial behavior they discover---either by manually searching themselves or through reports from community members. 
This approach is labor-intensive, making it difficult to scale to the amount of content generated by modern day platforms, and can be \rredit{practically and emotionally} challenging for the human moderators \rredit{\cite{gillespie_custodians_2018,gillespie_expanding_2020}}.
These challenges have led to increasing interest in the use of algorithmic tools to (partially)  automate the process of content moderation. Much of the recent work in this direction has focused on applying machine learning to automatically detect antisocial comments \cite{nobata_abusive_2016,wulczyn_ex_2017}, a technique which has gone on to see use in industry through tools like the Perspective API.\footnote{\url{https://www.perspectiveapi.com}}

While such tools have the potential to alleviate the labor-intensiveness of content moderation, they do not address one crucial limitation: this entire process is fundamentally \textit{reactive}, consisting of responding to antisocial content that has \textit{already} been posted. Even the best reactive moderation can only come after damage has already been done---possibly exposing any number of users to incivility and distracting or preventing productive discussions. Despite the technical attention that has been given to developing automatic classifiers for antisocial behavior, there has been surprisingly little work done to understand how well these tools actually fit into a human moderation workflow~\cite{jhaver_human-machine_2019}: Is locating uncivil comments for the purpose of reactive moderation actually the sole task where human moderators need algorithmic support? 
Or, are there other aspects of the moderation workflow where an algorithmic tool can be of help? 

In this work, we start addressing these questions by studying the potential for algorithmic assistance in a different moderation paradigm---\textit{proactive moderation}---\rredit{through a case study of a collaborative setting: Wikipedia Talk Pages.}
\rredit{We focus on a specific proactive moderation strategy that prior work has identified as needing algorithmic support:
identifying and monitoring conversations that are at-risk of devolving into uncivil behavior, with the goal of intervening early to avoid derailment or to mitigate its effects in a timely manner \cite{jurgens_just_2019,seering_moderator_2019}.}
We interview moderators to understand their proactive moderation practices and identify their needs for algorithmic support.  To explore the feasibility of addressing these needs with the technology available today, we build a prototype tool for assisting proactive moderation and observe how moderators engage with it in the context of their existing workflow. \rredit{In our analysis we pay particular attention to the ethical issues that come to the fore, and discuss their implications for system design and future work.}

Concretely, \rredit{our case study aims to address the following primary research questions}:
\begin{enumerate}
\item (How) do moderators act proactively to prevent the derailment of discussions into uncivil behavior? 
\item (How) can an algorithmic tool assist moderators in their proactive moderation workflow?
\end{enumerate}
\rredit{Building on the answers to these questions, we identify concrete next steps towards further developing such assisting technology so that it can support moderators in their existing proactive moderation workflow. } 

\section{Background: The Landscape of Moderation}
\label{sec:background}

\begin{figure}[ht]
    \centering
    \includegraphics[width=0.8\textwidth]{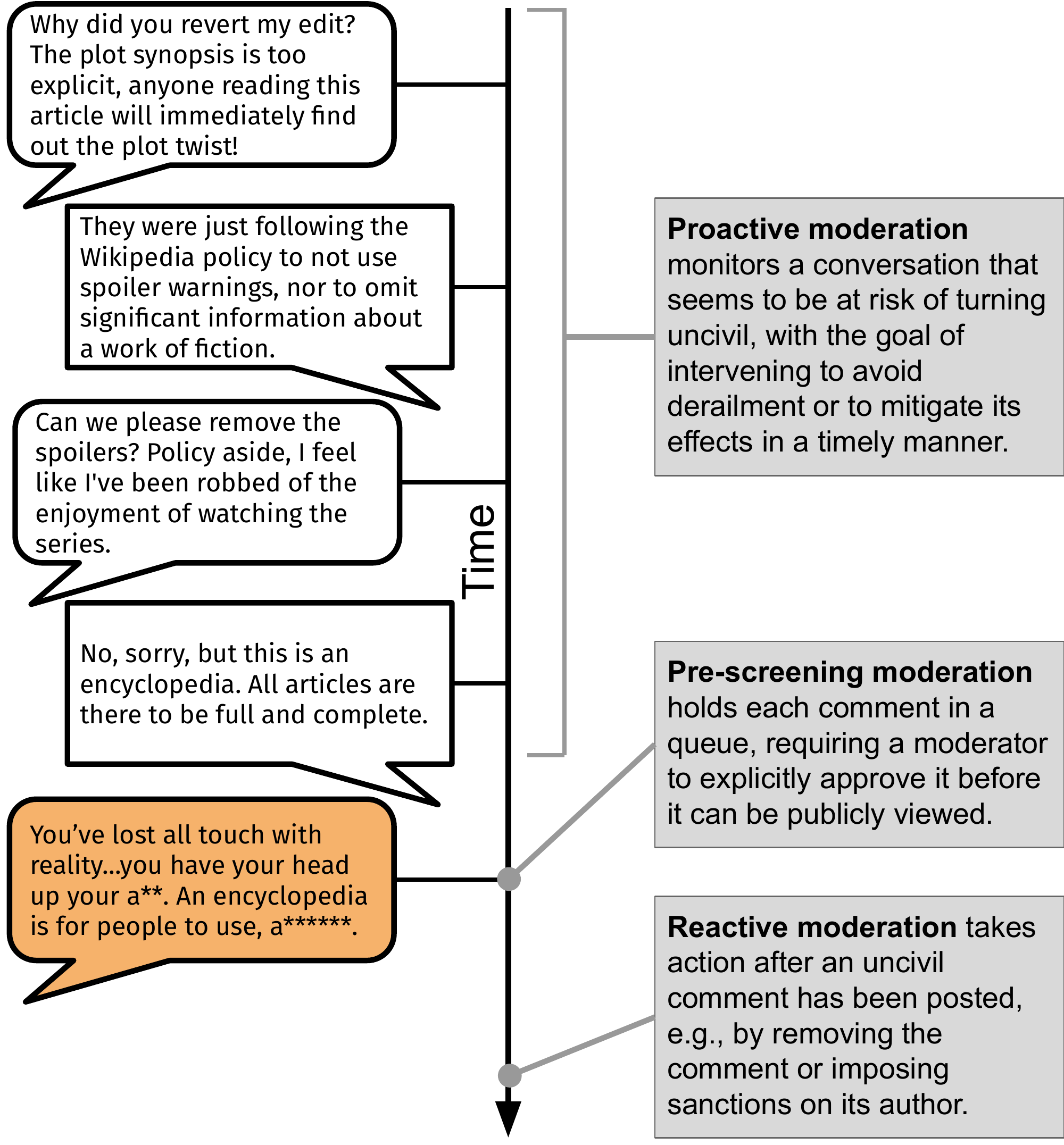}
    \caption{Three types of moderation paradigms exemplified in the context of a conversation between two Wikipedia editors that eventually derails into a personal attack (orange).}
    \label{fig:moderation_paradigms}
\end{figure}

\rredit{
Our work contributes to an extensive line of research on online discussion moderation which has aimed to both understand existing practices and develop new ones. Thus, to motivate our focus on the proactive paradigm and put our contributions in context, we begin by briefly surveying how prior work has characterized moderation practices, with an emphasis on identifying where existing practices fall on the reactive-proactive spectrum and exploring the role of algorithmic tools.
}

\rreditsec{\subsection{Practices of Moderation}}
\rredit{
In the popular consciousness, content moderation is often associated with the removal of illegal, hateful, or otherwise objectionable content \cite{gillespie_custodians_2018}, especially high-profile cases such as Reddit's 2015 mass banning of hate communities \cite{chandrasekharan_you_2017}. However, removal of content is but one part of content moderation, which in reality involves a wide and diverse range of hidden labor \cite{dosono_moderation_2019,wohn_volunteer_2019,matias_civic_2019} that, while less visible to the public eye than a high-profile ban, is no less important to maintaining the everyday functioning of an online community \cite{halfaker_rise_2013}.} 

\rredit{
The need to look beyond content removal is particularly salient in collaborative platforms like Wikipedia, which have eschewed top-down, platform-driven governance in favor of a community model of moderation \cite{seering_reconsidering_2020}, wherein ordinary community members volunteer to take on moderator roles \cite{halfaker_rise_2013,billings_understanding_2010}. Unlike platform-employed professional moderators, who are often seen as distant and separate from the communities they moderate \cite{seering_reconsidering_2020}, volunteer moderators are by definition part of their communities, and many will continue to participate in conversations and other informal interactions \cite{wohn_volunteer_2019,lo_when_2018}. Volunteer moderators must therefore balance their ``dual identities'' as both regular community members and authority figures. Different ways of managing this balance will result in different conceptions of one’s role and purpose as a moderator; in interviews, volunteer moderators have described their work with metaphors that range from the formal (``police'', ``governor'', ``manager'') to the informal (``team member'', ``facilitator'', ``adult in the room'')~\cite{seering_metaphors_2020}. This diversity in attitudes towards moderation naturally leads to a diversity in employed methodology. While many platforms that use volunteer moderation offer their moderators formal tools similar to those wielded by their platform-employed professional counterparts, such as the authority to remove comments \cite{gilbert_i_2020,jhaver_did_2019} or suspend users \cite{lo_when_2018,chang_trajectories_2019}, volunteer moderators often express a preference for softer, social approaches to keeping the community in line \cite{seering_moderator_2019}. Examples of such approaches include publicly modeling good behavior \cite{seering_shaping_2017}, educating users about the rules \cite{cai_what_2019}, and mediating disputes \cite{billings_understanding_2010}.
}

\rredit{
Given this wide breadth of moderation practices, researchers have sought to group and categorize them as a first step towards studying and understanding moderation \cite{seering_metaphors_2020,grimmelmann_virtues_2015}. In particular, an emerging line of work has proposed to distinguish different practices based on when they happen relative to the objectionable action being moderated: some practices are meant to respond to such behavior after it happens, while others are meant to discourage or prevent such behavior in the first place \cite{kiesler_regulating_2012}. The former group has been referred to as \emph{reactive} \cite{lo_when_2018,seering_moderator_2019} or \emph{ex-post} \cite{grimmelmann_virtues_2015} moderation, and the latter as \emph{proactive} or \emph{ex-ante}. Additionally, other work has identified a third set of practices in this space, known as \emph{pre-screening} moderation \cite{seering_metaphors_2020,kiesler_regulating_2012}, which occupy a middle ground by inserting an extra step of moderator approval in between the creation and public posting of user-created content.
In this work, we adopt the proactive/reactive/pre-screening categorization of moderation practices in the context of online discussion moderation. In the following sections (\ref{sec:reactive}-\ref{sec:proactive}) we discuss each in the context of online discussion platforms and compare their relative merits. Then, we survey the existing state of algorithmic approaches to moderation (\ref{sec:moderation_tools}) and identify a key gap in the literature, algorithmic approaches to proactive moderation, which we seek to fill.}

\subsection{Reactive Moderation  \rredit{of Online Discussions} }
\label{sec:reactive}
Currently, discussion moderation in online platforms most commonly takes the form of reacting to instances of antisocial behavior or rule violations~\cite{gillespie_custodians_2018}. The exact response to such content may vary~\cite{mcgillicuddy_controlling_2016}. On most large scale platforms, a standard response is to remove antisocial content that gets discovered by or reported to a centralized team of moderators~\cite{chandrasekharan_internets_2018,gillespie_custodians_2018}. Other platforms take a slightly softer approach and simply limit the visibility of such content rather than removing it outright~\cite{lampe_slashdot_2004}; some platforms extend this type of approach and limit content visibility on a \textit{user-specific} basis, allowing the blocking of content from known bad actors~\cite{jhaver_online_2018}. A further user-level response is to temporarily or permanently ban the authors of antisocial content from the platform~\cite{chang_trajectories_2019,jhaver_evaluating_2021}. 

Regardless of the exact nature of the response, however, reactive moderation in general is characterized by the fact that it relies on locating and responding to antisocial content that has \textit{already} been posted. Moderator responses in this paradigm thus serve as a form of post-hoc damage control, designed both to prevent the harm of the antisocial content from spreading too far, and to signal to the broader community that such behavior is unacceptable \cite{kiesler_regulating_2012}, thereby discouraging other users from reproducing it \cite{srinivasan_content_2019}.

While reactive moderation is widely used and studied, it also comes with an inherent weakness: because it involves taking action against antisocial content that has \textit{already} been posted, the offending content has an opportunity to be seen and to spread before moderators are able to take action, if they ever do at all. This can harm users exposed to the antisocial content, in addition to harming the platform by preventing or distracting from productive discussions.
For this reason, it has been argued that truly limiting the damage of antisocial behavior requires taking action before antisocial content can be seen by general audiences \cite{kiesler_regulating_2012}.

\rredit{In addition to potentially harming platforms and their users, reactive moderation strategies can further harm the \emph{moderators} tasked with regularly viewing and taking action on disturbing content, from the extreme stress of reacting to hateful comments to retribution from users in response to a moderation action threatening their personal safety \cite{dosono_moderation_2019, wohn_volunteer_2019, blackwell_when_2018}. We believe these harms implore us to investigate alternative approaches to moderation that may be able to avoid some of this human suffering.}

\subsection{Pre-screening Moderation \rredit{of Online Discussion}}
\label{sec:prescreening}
One approach to taking action before antisocial content can be seen by general audiences is \textit{pre-screening}: the paradigm by which all content must be reviewed and explicitly approved by moderators before it appears on the platform \rredit{\cite{kiesler_regulating_2012}}. This approach, hearkening back to the days of traditional pen-and-paper media, is still employed by a handful of platforms such as the \emph{New York Times} comment section.\footnote{\url{https://help.nytimes.com/hc/en-us/articles/115014792387-Comments}} Additionally, there has been work on hybrid automatic/human systems for comment pre-screening \cite{park_supporting_2016}. 

However, most platforms avoid this strategy because it raises a host of practical issues. Pre-screening is highly labor intensive, and scales poorly as a platform grows: for example, even with the help of algorithmic pre-screening, the \emph{New York Times} currently only allows comments on top stories for 8 hours during weekdays. Moreover, pre-screening prevents real-time interaction between users on a platform by introducing a delay between users submitting content and that content appearing on the platform while moderators review it. Finally, pre-screening has been subjected to criticism on the grounds of suppressing free speech \cite{gillespie_custodians_2018}. 

While pre-screening may prevent antisocial behavior from reaching general audiences, it still relies on moderators to react to and address all attempts at antisocial behavior users make on their platform---this reaction has just shifted from the public square to the moderators' private space. This shift limits the effectiveness of pre-screening by reproducing the problems described above.

\subsection{Proactive  Moderation \rredit{of Online Discussion}}
\label{sec:proactive}
\rredit{In online discussion platforms, proactive approaches to moderation can aim to discourage undesirable actions or to encourage prosocial behavior and productive conversations \cite{seering_shaping_2017,seering_moderator_2019}.  Proactive strategies can range from static design decisions to dynamic interventions in which moderators play a more active role.  They can also be broadly applied to the entire community or targeted towards specific situations, users, or conversations.}

\rredit{\emph{Static} strategies primarily involve deliberate choices in platform design aimed at promoting pro-social behaviors. These have a long and established history in social computing; now-common design choices such as activity indicators \cite{erickson_social_2000} and explicitly listed rules \cite{kiesler_regulating_2012} were initially developed as measures to encourage the development and adoption of social norms within online communities. More recent developments in this direction include limitations on community size or rate of participation \cite{grimmelmann_virtues_2015} and systems that prompt users to reflect more deeply on comments they have read \cite{kriplean_supporting_2012,kriplean_is_2012,jagannath_we_2020}. While such design elements have historically been broadly applied at the level of the entire platform, there has been work on UI-level interventions that are more targeted to specific situations \cite{halfaker_nice:_2011,lo_when_2018}. In the space of online discussions, such work has included psychological priming strategies that are deployed to users when they are about to comment in a discussion thread \cite{seering_designing_2019,taylor_accountability_2019}.}

\rredit{As platforms have grown and evolved, they have developed more \emph{dynamic} strategies for proactive community management, in which moderators take a more direct role. For instance, a natural development from static listing of rules involves moderators personally sending reminders of community rules in high-impact situations, such as when welcoming newcomers \cite{seering_moderator_2019}. As a further step from this, recent work has looked at how moderators can model good behavior in their own interactions, as a way of implicitly signaling to the community what proper behavior looks like~\cite{jagannath_we_2020,seering_shaping_2017}.}

\rredit{One dynamic strategy in particular has picked up increasing interest: moderators actively monitor ongoing conversations in order to proactively prevent them from derailing into uncivil behavior or, at least, to be in a position that allows them to mitigate the effects of derailment in a timely manner \cite{lo_when_2018, jurgens_just_2019}.  Unlike the more static strategies discussed above, this dynamic and highly-targeted moderation strategy requires substantial time and effort on behalf of the moderators and thus scales poorly. As such, prior work has advocated for offering algorithmic support for this type of strategy, in the form of ``predictive suggestions for threads or discussions that might soon devolve'' that could help ``moderators to focus their attention where it is needed.'' \cite{seering_moderator_2019}.  This current work aims to develop a system that can provide such algorithmic support and to test its feasibility.}

As with pre-screening moderation, proactively preventing conversations from derailing into incivility raises ethical concerns around censorship and free speech. We engage with these issues, assess the risk of potential misuses of a tool for assisting proactive moderation---such as shutting down conversations or blaming participants proactively based only on a prediction of future incivility---and discuss implications on its design. 
\rredit{While these concerns are warranted and deserving of a meaningful inquiry, we argue that the well-established harms to users, platforms, and moderators in the popular reactive moderation framework motivate us to explore alternate approaches to moderation. Because the status quo is not a neutral ethical choice, we must investigate such alternative approaches \emph{even though} it requires a careful ethical analysis. We return to these questions about the ethics of proactive moderation in Section~\ref{sec:ethics}}.

\rreditsec{\subsection{Algorithmic Tools for Moderation}\label{sec:moderation_tools}}
\rredit{Multiple studies on content moderation have identified a problem of scale: even if antisocial behavior is a small fraction of all content that gets posted, the sheer size of modern online platforms, together with the relatively small number of moderators present on most platforms, makes it infeasible for human moderators to keep up with all the content in need of moderation \cite{dosono_moderation_2019,lo_when_2018,wulczyn_ex_2017,gillespie_content_2020}. This has led to mental strain and burnout among moderators \cite{dosono_moderation_2019} and has directly inspired calls for the development of technological assistance to reduce the burden on human moderators \cite{wohn_volunteer_2019}. 
As Gillespie writes, ``the strongest argument for the automation of content moderation may be that, given the human costs, there is simply no other ethical way to do it, even if it is done poorly'' \cite{gillespie_content_2020}.
Technological responses to this call have ranged in complexity: basic tools include simple word-based filters \cite{wohn_volunteer_2019,lo_when_2018,chancellor_thyghgapp_2016} and blocklists \cite{jhaver_online_2018}, while more advanced systems attempt to use machine learning and natural language processing techniques to automatically identify antisocial content~\cite{wulczyn_ex_2017,nobata_abusive_2016,gamback_using_2017}.}

\rredit{Regardless of the choice of technical backend, most algorithmic tools for moderation are optimized for the reactive and pre-screening paradigms. A common use case is to apply the filter or classifier to content that has already been submitted for public posting; while in rare cases this can be applied as a pre-screening approach where the filter automatically blocks certain submitted content from getting posted (usually involving high-precision filters that look for hand-chosen terms known to cause problems in a specific micro-community) \cite{wohn_volunteer_2019,lo_when_2018}, the more common application is to use the filter or classifier to flag content for review by human moderators (allowing the content to stay public in the meantime) \cite{lo_when_2018,chandrasekharan_crossmod:_2019,jhaver_human-machine_2019,geiger_work_2010}, which results in a reactive moderation workflow.   }

\rredit{In comparison, algorithmic assistance for proactive moderation has been understudied. As described in Section \ref{sec:proactive}, research in the proactive space has largely focused on user-facing interventions rather than moderator-facing tools. Historically, this could be attributed to a technology gap: prior work has argued that an algorithmic tool for proactive moderation, with comparable scope to tools currently available for reactive and pre-screening moderation, would require technology that can predictively identify discussions that are about to devolve into antisocial behavior \cite{seering_moderator_2019}. Such technology has not been available until very recently, with a series of systems having been developed recently that are capable of making such ahead-of-time forecasts \cite{zhang_conversations_2018,liu_forecasting_2018,chang_trouble_2019}. We therefore identify an opportunity to explore the potential of such predictive technology to provide algorithmic assistance for proactive moderation. Our current work aims to explore this promising new direction.} %

\label{sec:related}
\rredit{
    \section{Case Study: Wikipedia Talk Pages}
    \label{sec:setting}
\rredit{In order to begin understanding the landscape of proactive moderation and potential for algorithmic assistance within the framework of community moderation, we conduct a case study of moderation on Wikipedia Talk Pages. On Wikipedia, discussions are not the primary content the platform provides; rather, Wikipedia hosts conversations on `Talk Pages' in order to facilitate discussion around the content of articles or broader policies governing the encyclopedia \cite{kittur_harnessing_2008}.\footnote{See the Wikipedia talk page guidelines: \url{https://en.wikipedia.org/wiki/Wikipedia:Talk_page}} In this collaborative, goal-driven discussion environment, antisocial behavior is particularly impactful, threatening the health of the editor community and disrupting productivity \cite{henner_wikimedia_2016,kittur_he_2007}.}

\rredit{Moderation of Talk Page discussions is community driven \cite{seering_reconsidering_2020}: the Wikipedia community elects administrators with broad technical powers on the platform such as deleting articles or blocking other users.\footnote{\url{https://en.wikipedia.org/wiki/Wikipedia:Administrators}} 
A subset of 
these administrators choose to engage in discussion moderation.
We note that there is no formal designation distinguishing discussion moderators from the rest of the administrators, and that discussion moderation practices (e.g., when a personal attack is subject for removal) are left largely at the discretion of these administrators.\footnote{In addition, the community grants an elected committee of arbitrators even broader powers to impose binding resolutions in order to resolve particularly severe disputes on Wikipedia, including but not limited to disputes in discussions (\url{https://en.wikipedia.org/wiki/Wikipedia:Arbitration}).} }

\rredit{The use of automated tools to assist in moderation has a long history on Wikipedia. Fully automated systems known as ``bots'' have been used to identify vandalism since as early as 2006~\cite{halfaker_rise_2013}, and to this day bots continue to play key roles on Wikipedia, not only in vandalism detection but also in more social aspects of community management such as welcoming and educating new users \cite{morgan_evaluating_2018,zheng_roles_2019}. While bots may be capable of handling mechanical tasks, other aspects of moderation still require a human touch, and for such tasks moderators make use of a different class of tools: ``assisted editing programs'' which are designed to augment (rather than replace) human moderation work \cite{geiger_work_2010}. A common design pattern in this space is to organize moderators' workload into work queues that help moderators prioritize situations in need of attention; this approach is exemplified by the popular tool Huggle \cite{geiger_work_2010} which organizes edits based on their algorithmically-determined likelihood of being vandalism. Beyond anti-vandalism, similar algorithmic approaches are used in another widely-used tool, ORES, to detect more types of rule violations in article edits \cite{halfaker_ores_2020}.} 

\rredit{Taken together, the goal-driven nature of Wikipedia Talk Page discussions, the large degree of discretion given to moderators, and familiarity Wikipedia moderators have with algorithmic tools in their workflow make this a convenient setting for an initial case study of proactive moderation practices. 
The fact that using automated tools is already a broadly accepted moderation practice on Wikipedia allows our study to focus specifically on the proactive aspect of our experimental tool rather than being confounded by moderators' thoughts on automation in general. The preponderance of existing tools on Wikipedia also gives us a good starting point from which to take design cues.
Moreover, we believe the goal-driven nature of the discussions provides moderators with a strong motivation to improve their moderation practices, while the large degree of discretion granted to Wikipedia moderators gives them the freedom to consider and attempt alternative strategies.}

\rredit{It is important to note upfront that the empirical setting for our case study does impose some limitations on our work. 
Given the unique structure and culture of Wikipedia,
our goal is not to report findings that generalize to any type of platform, but rather to begin understanding proactive moderation practices and the potential for algorithmic support in the specific setting of goal-driven online discussions. In the process, we provide a blueprint that other researchers can follow to begin understanding proactive moderation in other types of online communities, both for its own sake and for comparison with this setting.  } 

}

\section{Methods}
\label{sec:methods}

In order to begin to understand the moderators' proactive moderation workflow, as well as to investigate  the potential of integrating algorithmic assistance into this workflow, we engaged two methodological approaches: (1) we conducted interviews with moderators of public online discussions, and (2) developed a prototype tool for assisting proactive moderation. The interviews explored moderators'  experiences with proactive moderation in general, and also involved putting our prototype tool in their hands to observe how they may use a proactive moderation tool in practice and to inform its design. We engaged and iterated these approaches simultaneously, and each analysis was informed by the other.

\subsection{Interviews}
\label{sec:interviews}
\mredit{Following a rich line of prior literature that uses interviews to pull back the curtain on moderation practices \cite{gurzick_view_2009,dosono_moderation_2019, wohn_volunteer_2019, chandrasekharan_crossmod:_2019,seering_moderator_2019}, we conducted semi-structured interviews with nine administrators on Wikipedia who engage in Talk Page moderation.
Each interview was conducted over Zoom and lasted approximately one hour; we subsequently produced full de-identified transcripts and coded the data using thematic analysis.}

\mredit{The interviews had two primary goals. The first half of the interview focused on participants' current practices: the role of administrators in moderation on Wikipedia, the goals of moderation, the ways participants moderate proactively, and how they reason about the future of conversations to inform their proactive interventions. The second half of the interview focused on understanding the potential of an algorithmic tool for assisting proactive moderation.
By having participants use our prototype tool on real discussions happening live on Wikipedia Talk Pages, we had the opportunity to observe how it fits into their proactive moderation workflow, to what extent it addresses their needs, and to get feedback on design and ideal usage of such a tool directly from its target audience.
The generic script of the interviews is included in Appendix A.}

\mredit{We conducted these interviews with Institutional Review Board (IRB) approval and recruited participants through snowball sampling: by asking each participant to recommend any other individuals they know who do discussion moderation on Wikipedia. 
While our interviews provided invaluable direct access to moderators and their domain specific knowledge, this  recruitment procedure does impose a potential limitation on our work by potentially biasing our findings to the one branch of the Wikipedia moderator social graph that our sampling procedure reached.
}

\subsection{Prototype Tool for Assisting Proactive Moderation}
\begin{figure}[ht]
    \centering
    \includegraphics[width=0.75\textwidth]{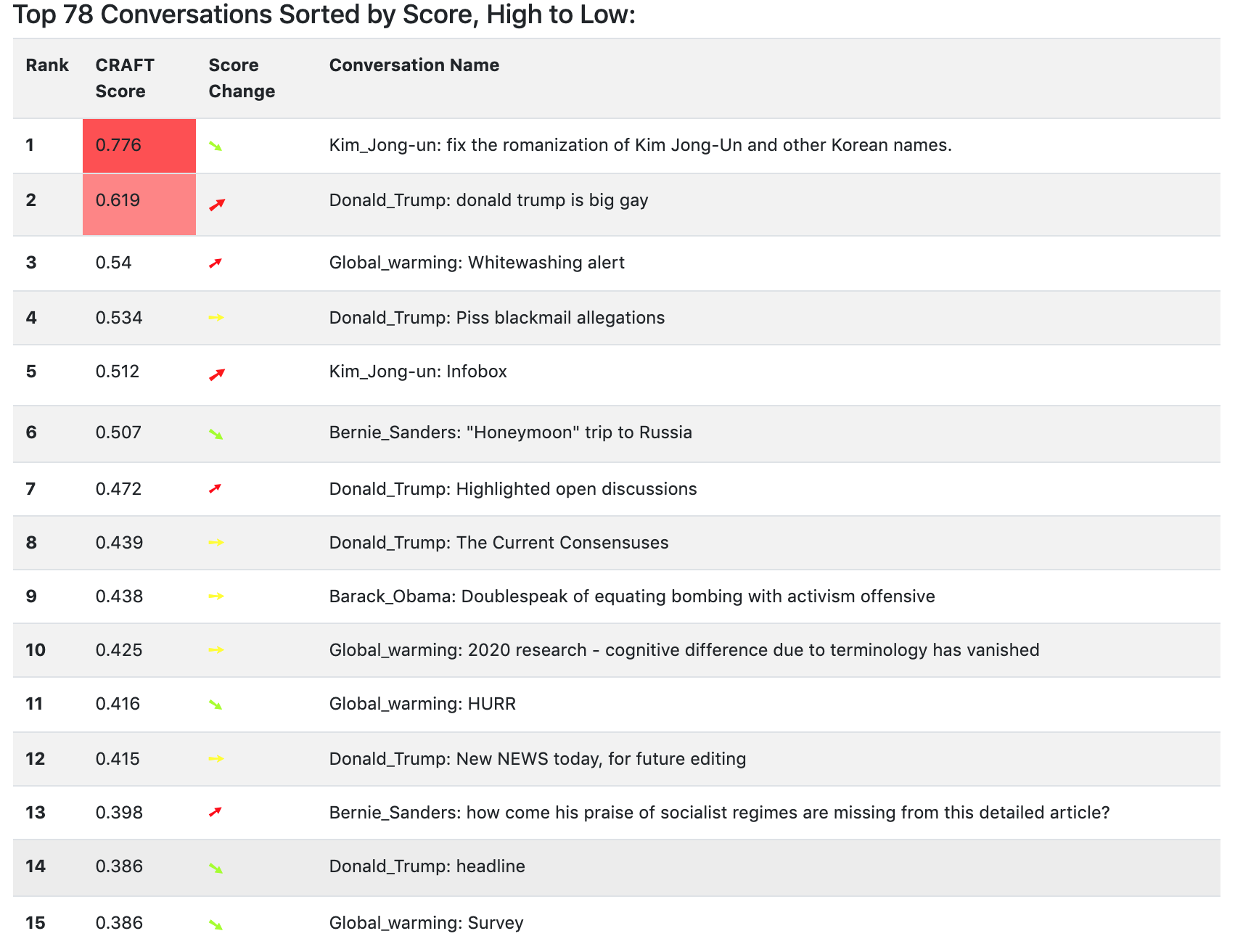}
    \caption{The \textit{Ranking View} of our prototype tool, showing a list of live conversations on Talk Pages, sorted by their predicted risk of derailing into antisocial behavior.}
    \label{fig:ranking_view}
\end{figure}

Our prototype tool  is implemented as a password protected website that includes two main features: a ranked view of ongoing conversations ordered according to their likelihood of derailing into future antisocial behavior (Figure \ref{fig:ranking_view}), and a conversation view giving a comment-by-comment breakdown of risk levels within the discussion (Figure \ref{fig:conversation_view}).  The tool currently works with discussions taking place on two different public discussion platforms  (Wikipedia Talk Pages and Reddit, although for the purpose of this work, we focus on the former) and is demonstrated in the Video Figure.\footnote{We use password protection to avoid any potential misuses of this technology.}$^,$\footnote{A link to the Video Figure can be found at \url{https://www.cs.cornell.edu/~cristian/Proactive_Moderation.html}.}  We now delve into the technical details of this tool.

\subsubsection{Backend: The CRAFT Architecture}
Our prototype tool is powered by a recent Natural Language Processing paradigm---conversational forecasting---which can be used to predict the future occurrence of an event in a conversation based on its current state \cite{zhang_conversations_2018,liu_forecasting_2018}.
\rredit{Prior work has applied this paradigm to the task of forecasting future antisocial behavior in online discussions, and found that algorithmic forecasting models are approaching the level of human intuition on this task, albeit with lower accuracy \cite{zhang_conversations_2018}. }
\rredit{While algorithmic models are not yet at the level of human performance, they \emph{are} at a level that is comparable to models already used in existing Wikipedia moderation tools: the current state-of-the-art model for forecasting future antisocial behavior, CRAFT, has a reported F1 score of 69.8\% \cite{chang_trouble_2019}, which is close to that of some models used in the popular ORES moderation tool.\footnote{\url{https://ores.wikimedia.org/v3/scores/enwiki/?model_info}}
We therefore believe that such models, while imperfect, are mature enough to be potentially useful to moderators.}
In light of this, we use the publicly available CRAFT model trained for forecasting future antisocial behavior on Wikipedia\footnote{Provided through the ConvoKit package (\url{https://convokit.cornell.edu}) \cite{chang_convokit_2020}.} to power our prototype tool.

Formally, given a discussion $D$, modeled as a sequence of $n$ comments in reply-order, $D = \lbrace c_1, c_2,...,c_n \rbrace$, CRAFT is trained to output the probability that a specified conversational event---in this case, antisocial behavior---will occur in the (not-yet-existent) next comment $c_{n+1}$ in $D$. In other words, $\text{CRAFT}(D)$ = $p(\text{isAntisocial}(c_{n+1}))$. CRAFT is an \textit{online} model, in that it updates its forecast for each new comment in the discussion. That is, once an actual comment $c_{n+1}$ happens, CRAFT can read the content of $c_{n+1}$ to produce an updated forecast $p(\text{isAntisocial}(c_{n+2}))$, and so on.

Since Wikipedia does not provide any special functionality to track live Talk Page conversations---in fact, Talk Pages are treated equivalently to normal article pages, with no special support for conversations---we implemented a system that parses and monitors changes in the discussions happening on a selected set of Wikipedia Talk pages\footnote{For this project, we choose pages that we reasoned are likely to have conflict and need moderation: Barack\_Obama, Bernie\_Sanders, Coronavirus\_disease\_2019, COVID-19\_pandemic, Donald\_Trump, Joe\_Biden, Kim\_Jong-un, and Global\_warming.} in real time.  To this end, we use the diff functionality from the Wikipedia API and use a set of heuristics to determine whether a change on the Talk Page represents a new comment in a discussion on that page, and which discussion it belongs to.  At regular intervals, the backend pulls the latest updates to every talk page being tracked, parses the updates to extract the conversations happening on the page, and runs CRAFT on those conversations to get an updated forecast of the risk of future incivility for each conversation. The tool also keeps track of how the CRAFT forecast for a discussion has changed over time. That is, for a (possibly ongoing) discussion $D = \lbrace c_1, c_2, c_3, \dots \rbrace$, the tool creates and maintains a history of CRAFT forecasts $\lbrace \text{CRAFT}(\lbrace c_1 \rbrace), \text{CRAFT}(\lbrace c_1, c_2 \rbrace), \text{CRAFT}(\lbrace c_1, c_2, c_3 \rbrace), \dots \rbrace$.

\subsubsection{Frontend: The Moderator's Display}
Our prototype frontend consists of two sections: a Ranking View and a Conversation View.  
\rredit{The frontend adopts 
design metaphors used in existing Wikipedia moderation tools, and comes with}
a broad range of features and parameters in order to engage interview participants in a discussion that can inform future design.

\rredit{On Wikipedia, assisted editing programs used by moderators often center around the organizational concept of the \emph{work queue}, in which content that was algorithmically determined to warrant human review is presented to moderators in a centralized, prioritized list \cite{halfaker_rise_2013,geiger_work_2010}. This is a proven design metaphor, having been used in empirically-studied moderator tools on Wikipedia \cite{geiger_work_2010} and elsewhere \cite{chandrasekharan_crossmod:_2019}. As such, our prototype tool similarly centers around a}
\textbf{Ranking View} (Figure \ref{fig:ranking_view}) 
as its main feature. Based on a list of Talk Pages to include, our prototype tool provides a CRAFT score ranking of all the conversations taking place on any of these pages, sorted in the order of their risk of derailing in the future. Each conversation is represented as a row in the ranking and is color coded to indicate its forecasted risk of derailing (shades of red, with darker red corresponding to higher risk). This reflects the most up to date CRAFT forecast for that conversation, computed based on all the comments posted so far in the conversation, i.e., $\text{CRAFT}(\lbrace c_1, c_2, \dots ,c_n \rbrace)$.  Additionally, to make it easy to identify rapidly escalating situations, each conversation in the ranking is also decorated with an arrow whose direction and size reflect the gradient and size of the most recent change in CRAFT forecast, i.e., $\text{CRAFT}(\lbrace c_1, c_2, \dots ,c_n \rbrace) - \text{CRAFT}(\lbrace c_1, c_2, \dots ,c_{n-1} \rbrace)$; for example a large red up-facing arrow would signal that the tension is rapidly rising in the respective conversation.

\begin{figure}[ht]
    \centering
    \includegraphics[width=0.75\textwidth]{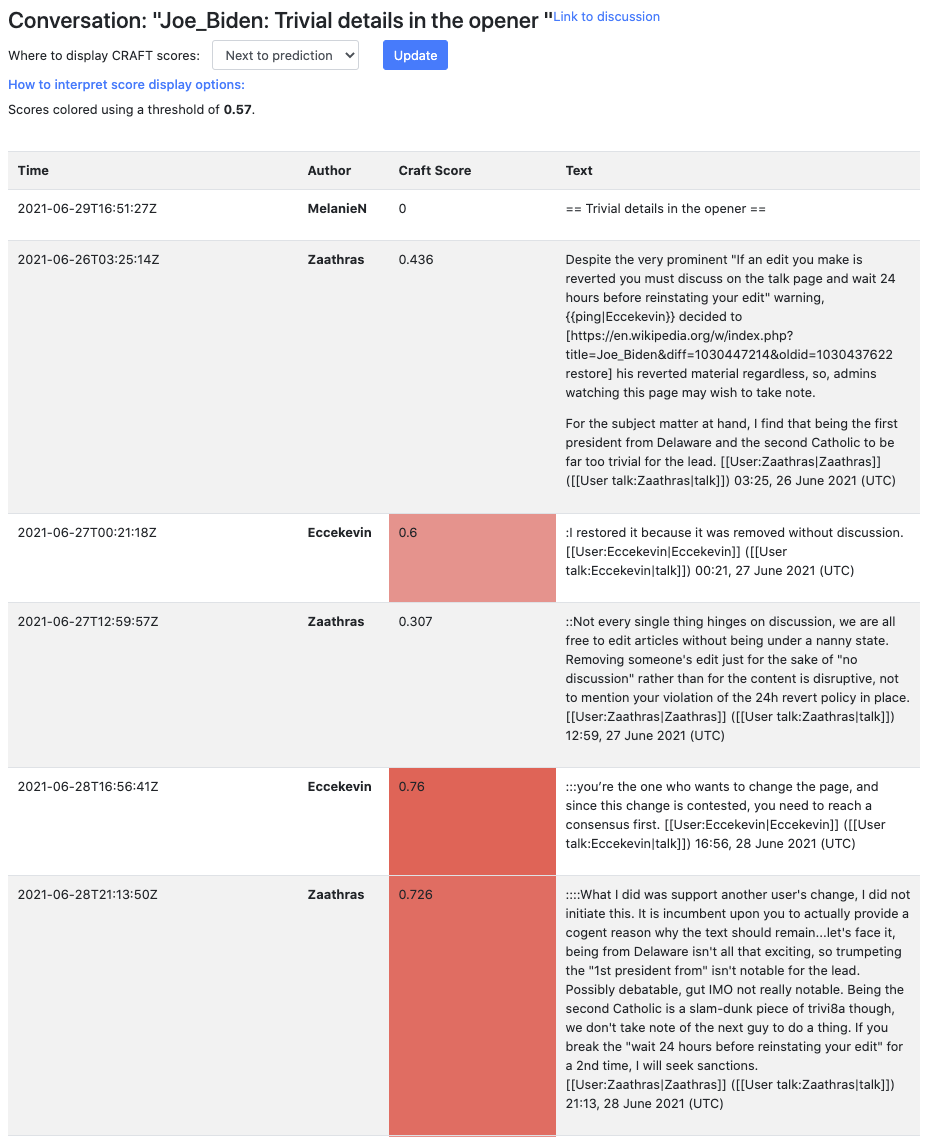}
    \caption{The \textit{Conversation View} of our prototype tool, showing a conversation with CRAFT scores alongside each comment. Each score represents the predicted risk of derailment at the time the corresponding comment was posted (taking into account the entire preceding context).}
    \label{fig:conversation_view}
\end{figure}

In addition to displaying summary level information about a conversation, each row of the Ranking view is a clickable link that leads to the 
\textbf{Conversation View} (Figure \ref{fig:conversation_view}), 
\mredit{which displays the entire history of that conversation.}
The Conversation View presents the text of each comment in the conversation along with the time it was posted, \mredit{its author,} 
 and the 
CRAFT score (color coded as before) at the time that comment was posted, 
i.e.,
taking into account the conversation up to and including that comment. This 
provides some level of transparency as to why the algorithm placed the conversation at a certain position in the Ranking View, allowing the moderator to observe how the predicted risk evolves as a conversation progresses.
\rredit{This design bears similarities to how algorithm decisions are presented to moderators in the experimental Reddit reactive moderation tool Crossmod \cite{chandrasekharan_crossmod:_2019}.}

\section{Findings}
\label{sec:findings}

\subsection{Moderator Goals: Content and Environment}
\label{sec:moderator_goals}
To contextualize our discussion, we start by understanding the broad goals moderators have in our particular domain of Wikipedia Talk Page discussions.  Following from the goal oriented nature of these discussions, as discussed in Section \ref{sec:setting},  participants highlighted how maintaining civil and productive discussions is not the end goal of their moderation. Rather, keeping discussions civil and functional is a crucial intermediary goal towards their primary goals: maintaining high quality content on the platform—in this case, encyclopedia articles—and maintaining a good environment for editors. As \textbf{P6} puts it:
\begin{quote}
    \textbf{P6}: [When I find a conversation headed downhill] I would not really care about the threads as having the thing go on, I'd care about the article and the environment of Wikipedia. I think those are the two things that I care about.
\end{quote}
Discussion moderation is crucial to maintaining these goals: antisocial behavior in discussions contributes directly to a hostile platform environment. Moreover, it can threaten the platform’s content when it pushes editors to give up on editing an article or leave Wikipedia altogether \cite{wikimedia_support_and_safety_team_harassment_2015}, or when it prevents or distracts from the conversations necessary for content creation and refinement. 
\rredit{This finding corroborates prior work showing how volunteer moderators are motivated by a variety of goals, including supporting a positive social environment and maintaining on-topic discussions relevant to their platform \cite{seering_reconsidering_2020, epstein_magic_2020, wohn_volunteer_2019}. }

A further consideration for moderators is that Wikipedia relies heavily on experienced users to contribute to the articles \rredit{\cite{halfaker_rise_2013}};
when these important content creators act uncivilly in a discussion, moderators are hesitant to sanction them because of their perceived value in writing articles \textit{even though} this incivility threatens the Wikipedia environment and alienates other users \cite{halfaker_dont_2011,collier_conflict_2012}. This exposes one way that the dual goals of moderation are in %
\rredit{tension on Wikipedia}. 
As \textbf{P3} explains:

\begin{quote}
    \textbf{P3}: I do believe that the English Wikipedia as a whole has a civility problem. [...] The community as a whole is far too willing to forgive incivility in the name of well—they're an experienced administrator or they're a really good content creator, so we'll just let them get by or say it wasn't that bad. And I think that that is not the path to a healthy community in the long term. I mean we have an editor retention problem, we know that. Everybody knows that. And I do think that the civility of the community is a significant part of that.
\end{quote}

In their view, moderators' imbalanced approach to the dual goals of moderation threatens the platform overall, and contributes to the difficulty retaining users on Wikipedia. \rredit{Any tool aimed at assisting moderators must consider both goals and not either in isolation.} 

\subsection{Proactive Moderation Practices}
\subsubsection{Acting Proactively}
\label{sec:acting_proactive}
Considering the broad goals of moderation on Wikipedia, we move to address one of our main research questions: Do moderators act proactively to prevent the derailment of discussions into antisocial behavior, and if so, what is their workflow?  

First, we confirm that moderators on Wikipedia do in fact engage in a variety of proactive moderation strategies.  The starting point in their workflow is their ability to foresee whether a conversation is at risk of derailing.  If they consider that this risk is elevated, they can further start monitoring it, or even decide to intervene in the discussion to avoid future derailment. For example:
\begin{quote}
    \textbf{P6}: Sometimes I can sit by and see things developing and I might drop by with a comment. {I don't tend to get involved in very big issues and charge in but I will go in and say, `This is becoming an inappropriate way of speaking. Let's talk collaboratively. Let's talk constructively.'}
    But do I monitor ongoing discussions for 
    it? 
    I suppose I look at some of the administrator notice boards, but I suppose I actually tend to sit more on the sidelines and watch other people engage in things, and only come in if I felt I had something to contribute or something to say like, `Tone this down.' And there is a good chance somebody else might too.
\end{quote}

While moderators' have access to formal administrative tools, called sanctions on Wikipedia\footnote{From the Wikipedia:Sanctions page: ``Sanctions are restrictions on editing Wikipedia that are applied to users or topic areas by the Wikipedia community and the Arbitration Committee in order to resolve disputes and curtail disruptive behaviour.'' (\url{https://en.wikipedia.org/wiki/Wikipedia:Sanctions})
}---such as blocking and interaction bans---proactively imposing any formal sanction is not permitted by Wikipedia's moderation guidelines and would raise ethical concerns; sanctions can only be used in response to a tangible offense.
Therefore, the proactive interventions that moderators can employ are limited to informal moderation techniques. 

Participants identify a variety of informal strategies they use to guide conversations which they assess to be at risk of derailment. For example, moderators will join a discussion as a level-headed participant in order to refocus the discussion on its original topic. \textbf{P5} explains their strategy:
\begin{quote}
    \textbf{P5}: In some of those cases I just engage as an additional participant rather than in discussion moderation just in order to just try and aid in those methods by bringing the discussion back on to context.
\end{quote}

A similar strategy is to leave just one comment in a discussion to acknowledge a growing dispute and try to neutralize it before it gets out of hand and irreparably damages the conversation. %
\rredit{Prior work has described this as a moderator acting as a ``mediator'', stepping into a conversation facing rising tensions in order to resolve conflicts between clashing discussion participants \cite{seering_metaphors_2020}. }
\textbf{P8}~explains their strategy:
\begin{quote}
    \textbf{P8}: I'll just leave a comment being like, `Hey guys, I think this might be going off topic,' and then I'll give my version of events. So it will be my opinion on it, in a very neutral way where I address each of their concerns. If I do it in a very polite way I think that typically a third party---especially an admin---does put the conversation back on topic.
\end{quote}
A different version of this strategy is to remind users of platform rules when moderators anticipate they will be violated. %
\rredit{Prior work has described this mode as a moderator acting as a ``referee'', working to ``resolve disputes by referencing rules or a body of accepted knowledge'' \cite{seering_metaphors_2020}. This can be seen as a more targeted version of automatic reminders, such as those triggered when interacting with newcomers \cite{halfaker_nice:_2011}.}
\textbf{P4} explains this strategy \rredit{as they apply it}:
\begin{quote} %
    \rredit{\textbf{P4}: 
    When we have [discussions in which there seems to be a significant chance of undesirable behavior arising], periodically we'll put up notices like, `Hey, remember to keep civil, keep your comments about the content of the discussion, not the other editors directly.'}
\end{quote}
These three interventions show the wide range in the depth of moderator involvement required for different proactive interventions. Joining a discussion as a participant to try to bring it back on track requires contextual knowledge of the conversation topic at hand and continued involvement in a discussion. Similarly, leaving one comment to address the concerns of discussion participants requires contextual knowledge of the conversation and topic at hand, but does not require ongoing engagement. Finally, reminding users of the platform's policies only requires a prediction of which policies may be violated, while the reminder itself can take the same form across discussions on different topics and does not require continued engagement. 

While some participants discuss how their proactive interventions can often bring discussions back on topic and avoid severe derailment beyond hope of repair, other participants describe how proactive interventions can backfire. Even when moderators forgo their formal sanctioning powers in favor of a softer approach, some users may react negatively to what they perceive as a threat of future sanctions. This implication may alienate users and limit the effectiveness of any proactive intervention. \textbf{P1} explains:
\begin{quote} %
    \textbf{P1}: I did [proactive interventions in discussions] much more when I was younger. It doesn’t work very well, I think because the idea is if you’re coming in as sort of like an uninvolved administrator, 
    \rredit{[...]}
    the assumed context is that you’re getting ready to sanction them, which is never as useful as a friendly reminder. 
    If I personally know one of the parties to the dispute, which happens on occasion, I might send them a direct email or a direct message,
    \rredit{[...]}
    just to try to hear what's going on. I found it particularly ineffective to post on Wiki to cool down, or something.
\end{quote}
This highlights one specific challenge moderators face when acting proactively: demonstrating to users that they genuinely want to help the conversation progress free of antisocial behavior rather than arriving early in preparation for future sanctions. 
\rredit{This corroborates prior work showing how discussion moderators may shy away from joining discussions despite a desire to do so, because of their role as a moderator \cite{gurzick_view_2009}. }
Thus, executing a successful proactive intervention requires a nuanced approach that considers the ways a moderator’s actions will be perceived by users.

\subsubsection{Benefits of Acting Proactively}
\label{sec:proactive_benefits}
In addition to the
\rredit{established} drawbacks of reactive moderation---and the respective benefits of the proactive paradigm---discussed in prior work and elaborated in Section \ref{sec:background}, our interviews reveal an additional issue: reactive interventions struggle to balance the two goals of moderation we identified above (Section \ref{sec:moderator_goals}), high-quality content creation and positive interactional environment. 

Since the reactive paradigm is only to act after a clear violation of community norms, in this case moderators can and do impose alienating formal sanctions. So, when experienced users who make otherwise valuable content behave antisocially, actions to sanction them---intended to maintain a healthy environment on the platform---alienate them and hence threaten the further development of the platform. On the other hand, protecting these antisocial users just because they create good content can cause disruption to the platform environment and alienate other users. \textbf{P7} explains this conundrum:
\begin{quote} %
    \textbf{P7}: 
    \rredit{[When experienced editors clash,]}
    that's where we, as administrators, sometimes have a very difficult task. We don't want to block experienced editors because they are very useful, very valuable.
    \rredit{[...]} By the same token, we don't want disruption. So, we've walked this very fine line where we try to hold experienced users who are sometimes misbehaving accountable without trying to block. It is a very difficult and fine line to walk and I think it would be nice if we had some way to better keep people civil, and better 
    \rredit{[...]}
    get people to work together. 
\end{quote}

Thus, in the reactive paradigm, antisocial content can threaten moderators' goals regardless of whether or not is addressed---disrupting the environment if it is not sanctioned, or alienating high value users if it is. Moreover, moderators face a significant challenge in realizing 
\rredit{their dual moderation goals}
in the face of incivility from established users through the reactive paradigm, threatening their emotional health and consuming a lot of their time. \textbf{P2} explains:
\begin{quote}
    \textbf{P2}: [When] someone has been incredibly uncivil to lots and lots of people, but he's also an incredibly influential editor, it is an excruciating process to kind of get through the kind of pieces that I need to to try and rein in his incivility. I just have to be patient, [because] it's ongoing and long.
\end{quote}
Therefore, addressing incivility from valuable content creators through the reactive paradigm threatens moderators themselves, in addition to their goals.  

Where reactive moderation faces this dilemma, the proactive paradigm offers a solution. Because proactive interventions come before any tangible incivility in a conversation, they are more well suited to take a softer and less alienating form, as discussed in Section \ref{sec:proactive_benefits}. This allows moderators to support a healthy environment by preventing incivility in discussions while avoiding the drawbacks of reactive strategies. \textbf{P2} explains their preference for using the proactive paradigm to address rising tensions in a conversation:
\begin{quote}
    \textbf{P2}: I did not become an administrator in order to block people. There are definitely people that became administrators because that's what they want to do, they wanted to police behavior. I actually spend a fair amount of time policing behavior in terms of my overall workload, but like I said, I try to operate in the social sphere and really kind of have conversations rather than using that.
\end{quote}
While not all moderators share this preference, proactive moderation offers those who do use it a more nuanced approach to moderation, better suited to balance 
\rredit{their multiple moderation goals}, rather than appeal directly to one or the other.

\subsubsection{Foreseeing Future Derailment}
\label{sec:foreseeing}
One crucial prerequisite of proactive moderation is identifying which conversations are at risk of derailing into incivility. We find that moderators on Wikipedia use their own intuition about the future trajectory of conversations towards this end, considering a variety of factors to internally reason about the future of the conversations they see. For example:
\begin{quote}
    Q: Given a civil conversation, do you think it is possible to predict if it will eventually derail into uncivil comments?
    
    \textbf{P7}: Yes. Not always but yes. I would say, certainly with experience, you get a feel for it where if a discussion has started off on the wrong foot, maybe someone got [their edits] reverted and then they opened, you know, maybe not an uncivil but kind of a terse message like, ``Hey, why did you undo my edit?,'' that's not uncivil but...It started things off on a bit of a wrong foot. I could guess that some of those things might get uncivil.
\end{quote}

Moderators use a variety of factors to make predictions about the future of conversations. Five participants report using direct observations from the conversation, like the conversation content or tone, to do forecasting. Using these direct features allows moderators to update their predictions over time as the conversation develops, whenever they check in on the conversation. 
On the other hand, the other four participants report forecasting solely based on metadata, including \mredit{features} of the conversation and of the \mredit{interlocutors.}
Salient conversation \mredit{properties} identified by participants include the ostensible conversation topic (as indicated by the conversation header) and the number of participants in the conversation. 
\mredit{Salient interlocutor metadata}  include level of experience on the platform, identity, and usernames.  Drawing on their past experiences, participants consider such features to estimate the risk that a conversation is likely to derail in the future.

\subsection{Evaluating the Feasibility of Algorithmic Assistance}
Equipped with an understanding of moderators' goals and practices, we now proceed to explore concrete ways in which an algorithmic tool can assist with their proactive moderation workflow.  We consider components of the workflow where moderators suggest that technical support is needed, and assess the feasibility of offering this support algorithmically with existing technology in an ethical and efficient manner.  We show the feasibility of algorithmic assistance with one crucial aspect of the workflow---discovering at risk conversations---and discuss design and ethical implications that arise from observing how moderators engage with our prototype tool.   We also discuss other needs that emerge from our interviews, but that do not lend themselves to algorithmic assistance due to both technical and ethical challenges, such as assigning blame for derailment or profiling users based on their prior behavior.

\subsubsection{Discovery of At-risk Conversations: Need and Support}
\label{sec:discovery_support}
We previously uncovered how moderators use their own intuition to decide which conversations to proactively moderate; now, we turn to the challenges moderators face in this crucial process and the resulting need for additional support.

\mredit{One idealized form of proactive moderation that all participants found appealing is to identify conversations that they suspect are highly likely to derail and monitor them so that they can intervene proactively at an opportune moment or to react immediately to any uncivil behavior that does arise.}
However, moderators' ability to identify at-risk conversations to monitor is limited by the scale of the platform. \textbf{P9} explains how even within the subset of topics they are interested in and engage in, their ability to \rredit{effectively} proactively monitor conversations is limited by their sheer number,
\rredit{which forces them to use only simplistic strategies, such as random discovery, to identify at-risk conversations to monitor}:
\begin{quote} %
    \textbf{P9}: There are too many [\rredit{potentially} at-risk conversations] to proactively monitor. I know there’s about 65 or 60 ongoing ones which are certainly always going to be at risk. \rredit{[...] So I usually either wait until I’m asked, or I happen to see something, or I skip around and happen to notice something.} %
\end{quote}

\rredit{The problem of scale is exacerbated by the inherent difficulty of determining when a conversation is in need of a proactive intervention. While \emph{every} participant we interviewed believes there are some contexts in which they can foresee derailment, as described in Section \ref{sec:foreseeing}, there is a wide range in how broad this context is and how confident participants are in their forecasts. Four participants believe that they can confidently forecast antisocial behavior in any Wikipedia context, but four others believe that they can only do so in very specific contexts with low confidence, and the last participant believes they can only make such forecasts in conversations on a handful of specific topics among discussion participants they know personally.}

\rredit{Given that moderators are often uncertain about their forecasts of a conversation's future trajectory, they may hesitate to intervene immediately, and instead desire to keep an eye on the situation to see how it develops. \textbf{P3} explains:
\begin{quote}
    \textbf{P3}: From time to time I do see a discussion I think that I want to monitor, and I'm like `Yeah, I probably should be keeping an eye on this.' [\dots] I might leave a tab open on it and come back to it just in case.
\end{quote}
As \textbf{P3} goes on to elaborate, however, this idealized notion of monitoring a conversation as it develops in real time is impractical in reality:
\begin{quote}
    \textbf{P3}: There are some technical challenges to [monitoring a discussion] just because of the way the Wikipedia software works. There isn't an easy way to say, `Give me updates for any changes in this discussion.' And, in fact, you can't even say, `Give me an update every time this page is changed,' which is a perennial source of annoyance. 
\end{quote} 
But on the other hand, the resulting gap in attention could cause the moderator to miss out on key developments in the conversation, and thereby lose an opportunity to intervene. \textbf{P6} explains this dilemma:}
\begin{quote}
    \textbf{P6}: I think I am okay at gauging if things are going to go pear-shaped, but do I always stick around to even find out if I am not interested in the topic? I may just move on and it blows up behind me. The hand grenade has gone off and I didn't even hear it because I've gone down the street. 
\end{quote}

\rredit{We therefore find that proactive moderation practices are difficult to scale up manually, both because of the size of the platform itself and because monitoring conversations---a necessary step given the uncertainty of moderators' forecasts---is time-consuming and impractical.}
Algorithmic solutions have the potential to address both challenges: by using a forecasting algorithm \rredit{rather than random discovery or other limited methods} to more efficiently identify conversations that are at risk, and by \rredit{automatically} monitoring them for relevant changes \rredit{as opposed to requiring manual, repeated checks.} 
This can potentially help moderators engage in proactive monitoring at a larger scale and dedicate more time to addressing potential issues.

\rredit{\textbf{How our prototype tool can address moderators' needs.}}  
\rredit{Having identified the potential ways an algorithmic tool could help scale up the process of proactive discussion moderation, we now investigate the extent to which our prototype tool meets this potential, as well as aspects in which it might still fall short and thereby provide directions for future work.} 
Concretely, we analyze moderators' feedback from their hands-on interaction with our prototype tool, with a focus on understanding which features moderators found the most useful and why, as well as what moderators found \textit{lacking} in the prototype tool.

Moderators' feedback on the prototype tool suggests that information presented in the tool's \textit{Ranking View} is helpful in discovering at-risk conversations, although individual moderators differed in their evaluation of exactly \textit{which} pieces of information were most useful. For example, \textbf{P4} reported that they would mainly use the CRAFT score to decide which conversations were worth monitoring:  
\begin{quote}
    \textbf{P4}: [For monitoring] I would just pick the ones with the highest score 'cause it seems to be somewhat accurate.
\end{quote}
Meanwhile, other participants highlighted the score change representation (i.e., the colored arrows) as providing an easy way to get a sense of when a monitored conversation needs to be further inspected. \textbf{P7} reports: 
\begin{quote}
    \textbf{P7}: I like the score change indicator. That is useful. From a cursory glance, it looks like if the score is going up, I would inspect it, if the score was going down, maybe it is not worth inspecting.
\end{quote}
All together, five participants described how both the score and score change representation would be useful towards discovering these at-risk conversations. 

However, moderators also identified several aspects of conversations that play into their existing intuitions about whether to monitor a conversation, but are not captured by the prototype tool. In particular, three participants mentioned wanting to know the ages of conversations, since their intuition is that if a conversation has been inactive for an extended period of time, it is unlikely to pick up again at all, much less turn uncivil. \textbf{P7} expresses this view:
\begin{quote}
    \textbf{P7}: That is very useful. That is probably all I would really need, too, except for the age of the conversation would also be another useful column because if the conversation was old, it wouldn't be worth my time to investigate it anyway but if I see the last comment was added within a day or two, I would then check it out. If it was more than, like, 2 or 3 days old, I mean, the conversation is probably dead anyway so it is not worth my time.
\end{quote}
Additionally, three participants wanted to see a summary of the end of the conversation---calling attention to the way that conversations on a path to incivility often stray far from the ostensible conversation topic, and how knowing the actual topic of discussion at hand is crucial for moderators to plan interventions. Both these suggestions could be taken into consideration in a future iteration of this tool.\footnote{It should be noted that the prototype tool already only includes “live” conversations and thus probably excludes most, if not all, “dead” conversations from the ranking, a fact that is not made apparent to the moderators. This issue notwithstanding, it is still possible that information about the age of conversations would still be useful to moderators even for more recently active conversations, so such a feature is still worth considering for future development.} On the other hand, five participants reported wanting to see data about discussion participants such as their usernames or age on the platform or their prior activity---features that could raise moral concerns and whose inclusion should thus be carefully weighed, as we will discuss in Section \ref{sec:ethics}.

The feedback discussed thus far
suggests that moderators would find the Ranking View useful in 
\emph{identifying} conversations that might be at risk. However, as discussed above, an important additional part of the proactive moderation workflow is continuing to \emph{monitor} such conversations---such that the ``grenade'' does not 
``blow up'' behind the moderator, to follow \textbf{P6}'s metaphor. %
While we believe the comment-by-comment information given by the Conversation View could be helpful for this,\footnote{In addition to just providing an augmented interface to follow the unfolding conversation, in a future iteration of the tool we can envision additional affordances, such as  allowing the moderator to request notifications based on specific CRAFT thresholds.} that would only be the case if this information aligns with how the moderator would intuitively judge the conversation.

\mredit{To assess this, we selected several conversations from different positions in the ranking and invited the moderators to first examine them raw (i.e., without added information), allowing them to make intuitive judgments, and then to re-examine them in the Conversation View.}
Overall, moderators reported that the displayed per-comment CRAFT scores matched their own 
\mredit{initial intuitive judgments}.  
For instance, while looking at an example conversation predicted to be heading for future incivility, \textbf{P2} describes:
\begin{quote}
    \textbf{P2}: [The escalating comment] definitely took it to a whole new level—and then having the third person come in, right? So, I feel like [the conversation view] is backing up what intuitively I had said. [...] I feel like that's very much in line with my experience and makes a lot of sense.
\end{quote}

The most notable exception is that some participants disagreed with the final CRAFT score of a conversation because they thought the conversation was unlikely to continue, and thus in a trivial sense unlikely to see any future incivility. \textbf{P8} explains:
\begin{quote}
    \textbf{P8}: I didn't think [the last comment has] that high [chance of seeing an antisocial reply]. I mean, in most cases this person [...] will rage quit. That's typically in my experience what happened. That's interesting. I didn't think it was going to be that high [of a score].
\end{quote}
While the prototype tool may be correctly capturing the intuition that the conversation is very tense and likely to see future antisocial behavior, in practice this may result in a user leaving the platform rather than leaving an antisocial comment in the conversation. Given that editor retention is also an issue of concern to Wikipedia moderators \cite{collier_conflict_2012}, this finding suggests that a future iteration of the tool could focus on more outcomes than just antisocial comments, as these other outcomes are equally part of the proactive moderation workflow.

Taking the design implications of a proactive moderation tool gleaned from this feedback together with the fact that the conversational forecasting algorithm underlying our prototype proactive moderation tool generally agrees with moderators' intuitions, we conclude that it is at least feasible to support moderators in identifying and monitoring at-risk conversations. However, this conclusion does not necessarily imply that moderators would accept and use such a tool, nor does it guarantee that the tool would be used in an ethical way. \textbf{P3} explains some hesitations:
\begin{quote}
    \textbf{P3}: I think an algorithm could be a useful indicator for flagging, `Hey, this seems like a topic or a conversation that might be a problem down the line.' But on its own I don't think an algorithm could actually be trusted to make the decision. A nice little browser plugin that highlights a section for me that says, `This discussion looks like it's getting heated, you might want to take a look at it,' that's something I would trust. A browser plug in telling me or a pile of machine learning telling me, `Block this person, they're making everything uncivil wherever they go,' not as inclined to trust it.
\end{quote}
As \textbf{P3} exemplifies, moderators are rightfully hesitant to put their full faith in an algorithmic tool, preferring to only use such a tool under their watch. Therefore, despite the agreement between state of the art forecasting methods and moderators' intuitions, these considerations motivate the need for follow-up work to conduct a large scale user study 
\rredit{to analyze the performance of the tool}
and analyze use patterns in a more systematic way. 

\subsubsection{Other Needs and Ethical Challenges}
\label{sec:ethics}
\mredit{Thus far we showed how our prototype tool addresses some proactive moderation needs, while leaving others unanswered. In the latter category, some gaps could be straightforwardly addressed by future iterations of the tool (e.g., considering the age of the conversation). Addressing others, however, could have significant ethical implications. We now consider such cases in more detail.}

In particular, one commonly reported complication in moderating antisocial behavior is how to properly assign \textit{blame} in multi-user interactions. \textbf{P8} points out that oftentimes, instances of antisocial behavior might arise in response to prior provocation:
\begin{quote}
    \textbf{P8}: [When considering taking moderation action] I have to read the whole [discussion] in order to understand where they're coming from. I'm looking to see how the other person [is] responding to them, because it's not really fair for me to stick to talk[ing] to one person if they’re being egged on. So, I tried to see where it got derailed and how to bring it back to a proper discussion.
\end{quote}
As P8 highlights, moderators do not want to treat a user being provoked and the user provoking them in the same way, and thus try to understand the root cause of a derailment in order to make a useful intervention and bring the discussion back on topic.
This need is heightened by the phenomenon of Civil POV Pushers---users who keep their language within community guidelines but come with an intention to only push their point of view, often enraging other users. While the term `Civil POV Pusher' is specific to Wikipedia, similar issues of bad-faith but superficially civil behavior could arise on any platform \cite{johnson_multiple_2017}. This makes it important for a moderator to know the history of a conversation to take action. \textbf{P3} explains:
\begin{quote}%
    \textbf{P3}: You need to look back: is somebody provoking them into this? Because that's always a possibility. 
    \rredit{[...]}
    We actually have a term kind of related to that. It's called civil POV pushing, for point of view, which is basically: you are here because you are on a mission. You want your opinion to be the correct one, but you are going to follow all the rules. You're not going to swear at people, but you are just going to keep bugging people or keep pressing your agenda until everybody else just gets tired of arguing with you. \rredit{[...]}  In that case it might not be justified that someone cussed them out, but it's more that they were provoked in a way. And so it probably not be as appropriate to take such a harsh sanction against the person who flew off the handle.
\end{quote}
As \textbf{P3} highlights, moderators use their discretion when sanctioning incivility if they believe the uncivil user was provoked by a bad-faith ``Civil POV pusher.'' This indicates the need to identify civil POV pushing in discussions, especially when it leads to that discussion derailing. 

In light of the complications introduced by the phenomenon of civil POV pushing, several moderators we spoke to expressed a desire for a further level of foresight in a proactive moderation tool: not just identifying \textit{which} conversations are at risk of derailing, but also \textit{why} those conversations might derail---even down to the level of exactly which individual \textit{comment} might trigger the derailment. Besides tying back to a general desire among moderators to address the root cause of antisocial behavior, such information could also have the practical benefit of helping identify civil POV pushers---a task that is currently acknowledged by Wikipedia as being one of the most challenging and patience-testing ones facing moderators.\footnote{\url{https://en.wikipedia.org/wiki/Wikipedia:Civil_POV_pushing}} 

However, we caution that algorithmic tools, at least in their current state, are not a good framework to support moderators in this need. The current forecasting models---including CRAFT, the one we use in our prototype tool---are based on statistical machine learning methods, which notably do \textit{not} explicitly model causality. In addition to being a key technical limitation of such models, this property can also introduce ethical hazards, as rather than making predictions based on the actual causal mechanisms, models might make predictions based on correlations arising from bias in the training data, a problem that has been previously observed in models for antisocial content detection \cite{wiegand_detection_2019,sap_risk_2019}. This limitation implies that, at least given current technologies, automated tools cannot be used to determine the root \textit{cause} of conversational derailment, and by extension cannot be used to distinguish the provoker from the provoked. Therefore, despite the depth of the challenge moderators face, algorithmic tools based on the statistical machine learning framework may not be an appropriate choice to assist moderators for this task given current limitations. 

These biases embedded into any moderation tool raise further questions about their use. For example, participants identified how they use information about the participants in a discussion to inform their perceptions of which conversations are at-risk of future incivility (Section \ref{sec:discovery_support}), and further that this use is a potential point for algorithmic support. However, automating the use of the identities of discussion participants for identifying at-risk conversations has questionable ethics. \textbf{P3} explains:
\begin{quote}
    \textbf{P3}: I think one thing that actually could be potentially useful for this is, though it also gets into some questionable territory is: who is in the discussion. Either just a breakdown of the top five contributors to the discussion. Or even, if we want to go into more Big Brother territory, [a summary of] how this person's comments usually score.
\end{quote}
The biases likely present within the underlying CRAFT conversational forecasting model, and any statistical machine learning model, suggest that profiling users based on how such an algorithm typically scores their comments is problematic, and may facilitate discrimination against \mredit{certain} users. 
\rredit{Moreover, because statistical machine learning algorithms' errors tend to be most concentrated around marginalized groups \cite{gillespie_content_2020, buolamwini_gender_2018}, profiling users based on their CRAFT scores over time holds the potential to reinforce \mredit{social inequities}
and further harm those already facing marginalization.}

Therefore, we caution against the deployment of any proactive moderation tools without implementing design features to discourage these fraught uses, in addition to informing moderators about the abilities and limitations of the tool. As mentioned in Section \ref{sec:discovery_support}, these dangers warrant closer investigation through a user study, to better understand how to best inform moderators about the tool's limitations and ethical hazards.

\section{Discussion}
\label{sec:discussion}
Motivated by the gap between the goals of moderation and the reality of reactive moderation strategies, this work seeks to deepen our understanding of the proactive moderation paradigm.   
\rredit{Through a case study of moderation on Wikipedia Talk Pages,
we uncover the workflow through which moderators proactively monitor and intervene in discussions they deem to be at risk of derailing into uncivil behavior.} 
We identify a specific component of this workflow that lends itself to algorithmic assistance and test the feasibility of such assistance by observing how moderators \rredit{in this community} engage with a tool running live on their platform.  

Based on our interviews with moderators of Wikipedia Talk Pages, 
we reveal a delicate balance between two moderation goals \rredit{in this collaborative setting}: maintaining a civil and welcoming environment, while trying not to alienate otherwise valuable content creators. 
Reactive moderation tends to put
these goals at odds: imposing harsh sanctions against uncivil behavior from otherwise valuable contributors can alienate them, but leaving such behavior alone creates a less civil and less welcoming environment.  
Proactive moderation offers an alternative pathway, by preventing sanctionable actions from occurring in the first place.  
Moreover, whereas reactive interventions tend to be strict formal sanctions such as a block, proactive interventions better lend themselves to more nuanced, informal actions. 
In interviews, moderators discuss how they employ proactive moderation strategies to prevent incivility without needing to remove any content or alienate \mrdelete{any} users.

The interviews further shed light on aspects of the proactive workflow that moderators currently find challenging, and which may therefore benefit from algorithmic assistance.  
\rredit{Moderators reported that there are too many ongoing conversations for them to reasonably inspect and that they would benefit from tools that can help them identify those that are at-risk, echoing suggestions from prior work \cite{seering_designing_2019,jurgens_just_2019}. Additionally, moderators indicated once at-risk conversations are identified, monitoring them is cumbersome and time consuming, and expressed a need for tools that can support this process.}

We 
\mredit{design a}
prototype tool to examine how algorithmic assistance could help moderators overcome these difficulties. In their feedback, moderators indicated that they found the prototype tool's \textit{Ranking View}, and in particular its CRAFT score and score change information, to be helpful in discovering potentially at-risk conversations. Furthermore, moderators' feedback on the \textit{Conversation View} shows that CRAFT scores within a conversation tend to match moderators' intuitions regarding derailment, \rredit{thus providing a helpful visual aid for monitoring at-risk conversations.}
Overall, these findings 
\rredit{suggest} that using an algorithmic tool to quickly identify candidate conversations for monitoring could therefore improve the scalability of proactive moderation by mitigating the need for labor-intensive manual searching and tracking of Talk Page discussions.  

\rredit{These findings motivate further steps towards developing and testing technology for assisting proactive moderation:}

\rredit{\textbf{Quantitative analysis.} While we have observed our prototype tool in the hands of moderators during the course of each hour long interview, a full quantitative analysis of the usage and utility of the tool would require observations of its use over a longer time period, running on a broad selection of discussions.  Such larger scale studies need to be set up with care, potentially using a sandbox approach \cite{chandrasekharan_crossmod:_2019}, avoiding the disruption of the broader Wikipedia environment. }

\rredit{\textbf{Potential misuses and ethical considerations.} Any new technology that engages so directly with human communication requires thorough ethical scrutiny.  
We have argued (Section \ref{sec:background}) that existing reactive and pre-screening moderation frameworks come with a number of ethical downsides, and that this justifies at least looking into alternative possibilities such as our proposal of algorithmically-assisted proactive moderation. Even so, we must remain conscious that our proposed methodology still falls under the broad umbrella of automated approaches to moderation, a subject area that in general raises hard moral questions \cite{gillespie_content_2020}. Though we have already identified and discussed some 
\mredit{ethical}
considerations that arose from the interviews (Section \ref{sec:ethics}), this is by no means comprehensive. At a high level, further moral questions may arise from two sources. The first is technological: while the flaws of toxicity-classification algorithms that underpin \emph{reactive} moderation tools are now well-studied and documented \cite{duarte_mixed_2018,davidson_automated_2017}, conversational forecasting algorithms like CRAFT are a much newer technology and have yet to receive similar levels of scrutiny. The second is social: though human moderators already engage in proactive moderation on a small scale, if tools like ours succeed in their goal of scaling up such practices, there may be unforeseen consequences, as past work has demonstrated in other situations where human moderators augmented their existing practices with automated tools \cite{halfaker_rise_2013}.}

The ethical considerations listed above provide 
\mredit{further inspiration for the design of a follow-up study.}
Such work should, in addition to evaluating the technical merits of algorithmically assisted proactive moderation, explore how to effectively inform moderators about the limitations and dangers of such algorithmic tools. Another important step 
is \textit{error analysis}: the proactive moderation paradigm is more prone to error than the reactive moderation paradigm, as predicting the future is inherently harder than detecting something that already happened. These errors can be more consequential in the proactive paradigm---a proactive intervention based on a false positive prediction could in the worst case shut down a perfectly civil conversation before it even has a chance to progress, with the potential for a chilling effect on speech if these errors are common. As such, future work needs to evaluate and characterize the types of errors made by forecasting algorithms, and observe how moderators might be influenced by these errors.  A particular focus should be placed on \textit{transparency} and \textit{explainability} to ensure that moderators can better understand the suggestions of the algorithms and identify erroneous ones.

\rredit{\textbf{Other domains and additional uses.} While our focus has been on the collaborative domain of Wikipedia Talk Pages, future work should investigate the proactive moderation paradigm in other settings, such as Reddit and Facebook Groups.  Even though our methodology can be easily translated to such domains, we do not expect the needs and dispositions of the moderators to be entirely echoed in other platforms, and thus the design of the tool might need to be adapted.  In particular, prior studies have revealed that Reddit moderators might be more hands off and thus less likely to engage in proactive strategies \cite{seering_moderator_2019}.}

Finally, our findings regarding how \textit{moderators} can benefit from predictions about the future of a conversation invite the question of what other ways this information can benefit the broader community. For example, future work could investigate the potential of empowering regular \textit{users}---that is, the \textit{participants} within online discussions---with this information through a user facing tool. By demonstrating the benefits of proactive moderation, and showing the feasibility of algorithmic assistance in this process, our present work lays a strong foundation for such future work, which constitutes a promising new direction in research on online content moderation.

\section*{Acknowledgements}

We would like to thank Lillian Lee, Cecelia Madsen, the 2021-2022 cohort of fellows at the Center for Advanced Study in the Behavioral Sciences at Stanford, and all the reviewers for the enlightening discussions and helpful suggestions. 
We additionally recognize everyone who helped with the implementation of the prototype tool, particularly Lucas Van Bramer and Oscar So for their contributions to the codebase, Todd Cullen for his help in setting up the backend server configuration, and Caleb Chiam, Liye Fu, Khonzoda Umarova, and Justine Zhang for their extensive testing and generous feedback. 
Finally, we are grateful to all the Wikipedia Administrators who participated in our interviews, as well as to Daniel Glus who guided our efforts by offering key starting insights into the Wikipedia moderation workflow, and who kickstarted the recruitment process by connecting us to the broader community.
This research was supported in part by an NSF CAREER award IIS-1750615; Jonathan P. Chang was supported in part by a fellowship with the Cornell Center for Social Sciences and Cristian Danescu-Niculescu-Mizil was supported in part by fellowships with the Cornell Center for Social Sciences and with the Center for Advanced Study in the Behavioral Sciences at Stanford.

\bibliographystyle{ACM-Reference-Format}
\bibliography{refs-cscw-moderation-final}


\begin{thebibliography}{67}


\ifx \showCODEN    \undefined \def \showCODEN     #1{\unskip}     \fi
\ifx \showDOI      \undefined \def \showDOI       #1{#1}\fi
\ifx \showISBNx    \undefined \def \showISBNx     #1{\unskip}     \fi
\ifx \showISBNxiii \undefined \def \showISBNxiii  #1{\unskip}     \fi
\ifx \showISSN     \undefined \def \showISSN      #1{\unskip}     \fi
\ifx \showLCCN     \undefined \def \showLCCN      #1{\unskip}     \fi
\ifx \shownote     \undefined \def \shownote      #1{#1}          \fi
\ifx \showarticletitle \undefined \def \showarticletitle #1{#1}   \fi
\ifx \showURL      \undefined \def \showURL       {\relax}        \fi
\providecommand\bibfield[2]{#2}
\providecommand\bibinfo[2]{#2}
\providecommand\natexlab[1]{#1}
\providecommand\showeprint[2][]{arXiv:#2}

\bibitem[\protect\citeauthoryear{Akbulut, Sahin, and Eristi}{Akbulut
  et~al\mbox{.}}{2010}]%
        {akbulut_cyberbullying_2010}
\bibfield{author}{\bibinfo{person}{Yavuz Akbulut},
  \bibinfo{person}{Yusuf~Levent Sahin}, {and} \bibinfo{person}{Bahadir
  Eristi}.} \bibinfo{year}{2010}\natexlab{}.
\newblock \showarticletitle{Cyberbullying {{Victimization}} among {{Turkish
  Online Social Utility Members}}}.
\newblock \bibinfo{journal}{\emph{Educational Technology \& Society}}
  \bibinfo{volume}{13}, \bibinfo{number}{4} (\bibinfo{date}{Oct.}
  \bibinfo{year}{2010}).
\newblock


\bibitem[\protect\citeauthoryear{Arazy, Yeo, and Nov}{Arazy
  et~al\mbox{.}}{2013}]%
        {arazy_stay_2013}
\bibfield{author}{\bibinfo{person}{Ofer Arazy}, \bibinfo{person}{Lisa Yeo},
  {and} \bibinfo{person}{Oded Nov}.} \bibinfo{year}{2013}\natexlab{}.
\newblock \showarticletitle{Stay on the {{Wikipedia Task}}: {{When Task-related
  Disagreements Slip Into Personal}} and {{Procedural Conflicts}}}.
\newblock \bibinfo{journal}{\emph{Journal of the American Society for
  Information Science and Technology}} \bibinfo{volume}{64},
  \bibinfo{number}{8} (\bibinfo{date}{Aug.} \bibinfo{year}{2013}).
\newblock


\bibitem[\protect\citeauthoryear{Billings and Watts}{Billings and
  Watts}{2010}]%
        {billings_understanding_2010}
\bibfield{author}{\bibinfo{person}{Matt Billings} {and}
  \bibinfo{person}{Leon~A. Watts}.} \bibinfo{year}{2010}\natexlab{}.
\newblock \showarticletitle{Understanding Dispute Resolution Online: Using Text
  to Reflect Personal and Substantive Issues in Conflict}. In
  \bibinfo{booktitle}{\emph{Proceedings of {{CHI}}}}.
\newblock


\bibitem[\protect\citeauthoryear{Blackwell, Chen, Schoenebeck, and
  Lampe}{Blackwell et~al\mbox{.}}{2018}]%
        {blackwell_when_2018}
\bibfield{author}{\bibinfo{person}{Lindsay Blackwell},
  \bibinfo{person}{Tianying Chen}, \bibinfo{person}{Sarita Schoenebeck}, {and}
  \bibinfo{person}{Cliff Lampe}.} \bibinfo{year}{2018}\natexlab{}.
\newblock \showarticletitle{When {{Online Harassment}} Is {{Perceived}} as
  {{Justified}}}. In \bibinfo{booktitle}{\emph{Proceedings of {{ICWSM}}}}.
\newblock


\bibitem[\protect\citeauthoryear{Buolamwini and Gebru}{Buolamwini and
  Gebru}{2018}]%
        {buolamwini_gender_2018}
\bibfield{author}{\bibinfo{person}{Joy Buolamwini} {and}
  \bibinfo{person}{Timnit Gebru}.} \bibinfo{year}{2018}\natexlab{}.
\newblock \showarticletitle{Gender {{Shades}}: {{Intersectional Accuracy
  Disparities}} in {{Commercial Gender Classification}}}. In
  \bibinfo{booktitle}{\emph{Proceedings of {{FAccT}}}}.
\newblock


\bibitem[\protect\citeauthoryear{Cai and Wohn}{Cai and Wohn}{2019}]%
        {cai_what_2019}
\bibfield{author}{\bibinfo{person}{Jie Cai} {and}
  \bibinfo{person}{Donghee~Yvette Wohn}.} \bibinfo{year}{2019}\natexlab{}.
\newblock \showarticletitle{What Are {{Effective Strategies}} of {{Handling
  Harassment}} on {{Twitch}}? {{Users}}' {{Perspectives}}}. In
  \bibinfo{booktitle}{\emph{Proceedings of {{CSCW}}}}.
\newblock


\bibitem[\protect\citeauthoryear{Chancellor, Pater, Clear, Gilbert, and
  De~Choudhury}{Chancellor et~al\mbox{.}}{2016}]%
        {chancellor_thyghgapp_2016}
\bibfield{author}{\bibinfo{person}{Stevie Chancellor},
  \bibinfo{person}{Jessica~Annette Pater}, \bibinfo{person}{Trustin Clear},
  \bibinfo{person}{Eric Gilbert}, {and} \bibinfo{person}{Munmun De~Choudhury}.}
  \bibinfo{year}{2016}\natexlab{}.
\newblock \showarticletitle{\#thyghgapp: {{Instagram Content Moderation}} and
  {{Lexical Variation}} in {{Pro-Eating Disorder Communities}}}. In
  \bibinfo{booktitle}{\emph{Proceedings of {{CSCW}}}}.
\newblock


\bibitem[\protect\citeauthoryear{Chandrasekharan, Gandhi, Mustelier, and
  Gilbert}{Chandrasekharan et~al\mbox{.}}{2019}]%
        {chandrasekharan_crossmod:_2019}
\bibfield{author}{\bibinfo{person}{Eshwar Chandrasekharan},
  \bibinfo{person}{Chaitrali Gandhi}, \bibinfo{person}{Matthew~Wortley
  Mustelier}, {and} \bibinfo{person}{Eric Gilbert}.}
  \bibinfo{year}{2019}\natexlab{}.
\newblock \showarticletitle{Crossmod: {{A Cross-Community Learning-based
  System}} to {{Assist Reddit Moderators}}}. In
  \bibinfo{booktitle}{\emph{Proceedings of {{CSCW}}}}.
\newblock


\bibitem[\protect\citeauthoryear{Chandrasekharan, Pavalanathan, Srinivasan,
  Glynn, Eisenstein, and Gilbert}{Chandrasekharan et~al\mbox{.}}{2017}]%
        {chandrasekharan_you_2017}
\bibfield{author}{\bibinfo{person}{Eshwar Chandrasekharan},
  \bibinfo{person}{Umashanthi Pavalanathan}, \bibinfo{person}{Anirudh
  Srinivasan}, \bibinfo{person}{Adam Glynn}, \bibinfo{person}{Jacob
  Eisenstein}, {and} \bibinfo{person}{Eric Gilbert}.}
  \bibinfo{year}{2017}\natexlab{}.
\newblock \showarticletitle{You {{Can}}'t {{Stay Here}}: {{The Efficacy}} of
  {{Reddit}}'s 2015 {{Ban Examined Through Hate Speech}}}. In
  \bibinfo{booktitle}{\emph{Proceedings of {{CSCW}}}}.
\newblock


\bibitem[\protect\citeauthoryear{Chandrasekharan, Samory, Jhaver, Charvat,
  Bruckman, Lampe, Eisenstein, and Gilbert}{Chandrasekharan
  et~al\mbox{.}}{2018}]%
        {chandrasekharan_internets_2018}
\bibfield{author}{\bibinfo{person}{Eshwar Chandrasekharan},
  \bibinfo{person}{Mattia Samory}, \bibinfo{person}{Shagun Jhaver},
  \bibinfo{person}{Hunter Charvat}, \bibinfo{person}{Amy Bruckman},
  \bibinfo{person}{Cliff Lampe}, \bibinfo{person}{Jacob Eisenstein}, {and}
  \bibinfo{person}{Eric Gilbert}.} \bibinfo{year}{2018}\natexlab{}.
\newblock \showarticletitle{The {{Internet}}'s {{Hidden Rules}}: {{An Empirical
  Study}} of {{Reddit Norm Violations}} at {{Micro}}, {{Meso}}, and {{Macro
  Scales}}}. In \bibinfo{booktitle}{\emph{Proceedings of {{CSCW}}}}.
\newblock


\bibitem[\protect\citeauthoryear{Chang, Chiam, Fu, Wang, Zhang, and
  {Danescu-Niculescu-Mizil}}{Chang et~al\mbox{.}}{2020}]%
        {chang_convokit_2020}
\bibfield{author}{\bibinfo{person}{Jonathan~P. Chang}, \bibinfo{person}{Caleb
  Chiam}, \bibinfo{person}{Liye Fu}, \bibinfo{person}{Andrew Wang},
  \bibinfo{person}{Justine Zhang}, {and} \bibinfo{person}{Cristian
  {Danescu-Niculescu-Mizil}}.} \bibinfo{year}{2020}\natexlab{}.
\newblock \showarticletitle{{{ConvoKit}}: {{A Toolkit}} for the {{Analysis}} of
  {{Conversations}}}. In \bibinfo{booktitle}{\emph{Proceedings of
  {{SIGDIAL}}}}.
\newblock


\bibitem[\protect\citeauthoryear{Chang and {Danescu-Niculescu-Mizil}}{Chang and
  {Danescu-Niculescu-Mizil}}{2019a}]%
        {chang_trajectories_2019}
\bibfield{author}{\bibinfo{person}{Jonathan~P. Chang} {and}
  \bibinfo{person}{Cristian {Danescu-Niculescu-Mizil}}.}
  \bibinfo{year}{2019}\natexlab{a}.
\newblock \showarticletitle{Trajectories of {{Blocked Community Members}}:
  {{Redemption}}, {{Recidivism}} and {{Departure}}}. In
  \bibinfo{booktitle}{\emph{Proceedings of {{WWW}}}}.
\newblock


\bibitem[\protect\citeauthoryear{Chang and {Danescu-Niculescu-Mizil}}{Chang and
  {Danescu-Niculescu-Mizil}}{2019b}]%
        {chang_trouble_2019}
\bibfield{author}{\bibinfo{person}{Jonathan~P. Chang} {and}
  \bibinfo{person}{Cristian {Danescu-Niculescu-Mizil}}.}
  \bibinfo{year}{2019}\natexlab{b}.
\newblock \showarticletitle{Trouble on the {{Horizon}}: {{Forecasting}} the
  {{Derailment}} of {{Online Conversations}} as They {{Develop}}}. In
  \bibinfo{booktitle}{\emph{Proceedings of {{EMNLP}}}}.
\newblock


\bibitem[\protect\citeauthoryear{Collier and Bear}{Collier and Bear}{2012}]%
        {collier_conflict_2012}
\bibfield{author}{\bibinfo{person}{Benjamin Collier} {and}
  \bibinfo{person}{Julia Bear}.} \bibinfo{year}{2012}\natexlab{}.
\newblock \showarticletitle{Conflict, {{Criticism}}, or {{Confidence}}: {{An
  Empirical Examination}} of the {{Gender Gap}} in {{Wikipedia
  Contributions}}}. In \bibinfo{booktitle}{\emph{Proceedings of {{CSCW}}}}.
\newblock


\bibitem[\protect\citeauthoryear{Davidson, Warmsley, Macy, and Weber}{Davidson
  et~al\mbox{.}}{2017}]%
        {davidson_automated_2017}
\bibfield{author}{\bibinfo{person}{Thomas Davidson}, \bibinfo{person}{Dana
  Warmsley}, \bibinfo{person}{Michael Macy}, {and} \bibinfo{person}{Ingmar
  Weber}.} \bibinfo{year}{2017}\natexlab{}.
\newblock \showarticletitle{Automated {{Hate Speech Detection}} and the
  {{Problem}} of {{Offensive Language}}}. In
  \bibinfo{booktitle}{\emph{Proceedings of {{ICWSM}}}}.
\newblock


\bibitem[\protect\citeauthoryear{Dosono and Semaan}{Dosono and Semaan}{2019}]%
        {dosono_moderation_2019}
\bibfield{author}{\bibinfo{person}{Bryan Dosono} {and} \bibinfo{person}{Bryan
  Semaan}.} \bibinfo{year}{2019}\natexlab{}.
\newblock \showarticletitle{Moderation {{Practices}} as {{Emotional Labor}} in
  {{Sustaining Online Communities}}: {{The Case}} of {{AAPI Identity Work}} on
  {{Reddit}}}. In \bibinfo{booktitle}{\emph{Proceedings of {{CHI}}}}.
\newblock


\bibitem[\protect\citeauthoryear{Duarte and Llans{\'o}}{Duarte and
  Llans{\'o}}{2018}]%
        {duarte_mixed_2018}
\bibfield{author}{\bibinfo{person}{Natasha Duarte} {and} \bibinfo{person}{Emma
  Llans{\'o}}.} \bibinfo{year}{2018}\natexlab{}.
\newblock \showarticletitle{Mixed {{Messages}}? {{The Limits}} of {{Automated
  Social Media Content Analysis}}}. In \bibinfo{booktitle}{\emph{Proceedings of
  {{FAccT}}}}.
\newblock


\bibitem[\protect\citeauthoryear{Epstein and Leshed}{Epstein and
  Leshed}{2020}]%
        {epstein_magic_2020}
\bibfield{author}{\bibinfo{person}{Dmitry Epstein} {and} \bibinfo{person}{Gilly
  Leshed}.} \bibinfo{year}{2020}\natexlab{}.
\newblock \showarticletitle{The {{Magic Sauce}}: {{Practices}} of
  {{Facilitation}} in {{Online Policy Deliberation}}}.
\newblock \bibinfo{journal}{\emph{Journal of Deliberative Democracy}}
  \bibinfo{volume}{12}, \bibinfo{number}{1} (\bibinfo{date}{May}
  \bibinfo{year}{2020}).
\newblock


\bibitem[\protect\citeauthoryear{Erickson and Kellogg}{Erickson and
  Kellogg}{2000}]%
        {erickson_social_2000}
\bibfield{author}{\bibinfo{person}{Thomas Erickson} {and}
  \bibinfo{person}{Wendy~A. Kellogg}.} \bibinfo{year}{2000}\natexlab{}.
\newblock \showarticletitle{Social Translucence: An Approach to Designing
  Systems That Support Social Processes}.
\newblock \bibinfo{journal}{\emph{ACM Transactions on Computer-Human
  Interaction}} \bibinfo{volume}{7}, \bibinfo{number}{1} (\bibinfo{date}{March}
  \bibinfo{year}{2000}).
\newblock


\bibitem[\protect\citeauthoryear{Gamb{\"a}ck and Sikdar}{Gamb{\"a}ck and
  Sikdar}{2017}]%
        {gamback_using_2017}
\bibfield{author}{\bibinfo{person}{Bj{\"o}rn Gamb{\"a}ck} {and}
  \bibinfo{person}{Utpal~Kumar Sikdar}.} \bibinfo{year}{2017}\natexlab{}.
\newblock \showarticletitle{Using {{Convolutional Neural Networks}} to
  {{Classify Hate-Speech}}}. In \bibinfo{booktitle}{\emph{Proceedings of
  {{ALW}}}}.
\newblock


\bibitem[\protect\citeauthoryear{Geiger and Ribes}{Geiger and Ribes}{2010}]%
        {geiger_work_2010}
\bibfield{author}{\bibinfo{person}{R.~Stuart Geiger} {and}
  \bibinfo{person}{David Ribes}.} \bibinfo{year}{2010}\natexlab{}.
\newblock \showarticletitle{The Work of Sustaining Order in Wikipedia: The
  Banning of a Vandal}. In \bibinfo{booktitle}{\emph{Proceedings of {{CSCW}}}}.
\newblock


\bibitem[\protect\citeauthoryear{Gilbert}{Gilbert}{2020}]%
        {gilbert_i_2020}
\bibfield{author}{\bibinfo{person}{Sarah~A. Gilbert}.}
  \bibinfo{year}{2020}\natexlab{}.
\newblock \showarticletitle{"{{I}} Run the World's Largest Historical Outreach
  Project and It's on a Cesspool of a Website." {{Moderating}} a {{Public
  Scholarship Site}} on {{Reddit}}: {{A Case Study}} of r/{{AskHistorians}}}.
  In \bibinfo{booktitle}{\emph{Proceedings of {{CSCW}}}}.
\newblock


\bibitem[\protect\citeauthoryear{Gillespie}{Gillespie}{2018}]%
        {gillespie_custodians_2018}
\bibfield{author}{\bibinfo{person}{Tarleton Gillespie}.}
  \bibinfo{year}{2018}\natexlab{}.
\newblock \bibinfo{booktitle}{\emph{Custodians of the {{Internet}}:
  {{Platforms}}, {{Content Moderation}}, and the {{Hidden Decisions}} That
  {{Shape Social Media}}}}.
\newblock \bibinfo{publisher}{{Yale University Press}}, \bibinfo{address}{{New
  Haven}}.
\newblock


\bibitem[\protect\citeauthoryear{Gillespie}{Gillespie}{2020}]%
        {gillespie_content_2020}
\bibfield{author}{\bibinfo{person}{Tarleton Gillespie}.}
  \bibinfo{year}{2020}\natexlab{}.
\newblock \showarticletitle{Content Moderation, {{AI}}, and the Question of
  Scale:}.
\newblock \bibinfo{journal}{\emph{Big Data \& Society}} (\bibinfo{date}{July}
  \bibinfo{year}{2020}).
\newblock


\bibitem[\protect\citeauthoryear{Gillespie, Aufderheide, Carmi, Gerrard, Gorwa,
  {Matamoros-Fern{\'a}ndez}, Roberts, Sinnreich, and Myers~West}{Gillespie
  et~al\mbox{.}}{2020}]%
        {gillespie_expanding_2020}
\bibfield{author}{\bibinfo{person}{Tarleton Gillespie},
  \bibinfo{person}{Patricia Aufderheide}, \bibinfo{person}{Elinor Carmi},
  \bibinfo{person}{Ysabel Gerrard}, \bibinfo{person}{Robert Gorwa},
  \bibinfo{person}{Ariadna {Matamoros-Fern{\'a}ndez}},
  \bibinfo{person}{Sarah~T. Roberts}, \bibinfo{person}{Aram Sinnreich}, {and}
  \bibinfo{person}{Sarah Myers~West}.} \bibinfo{year}{2020}\natexlab{}.
\newblock \showarticletitle{Expanding the Debate about Content Moderation:
  {{Scholarly}} Research Agendas for the Coming Policy Debates}.
\newblock \bibinfo{journal}{\emph{Internet Policy Review}} \bibinfo{volume}{9},
  \bibinfo{number}{4} (\bibinfo{date}{Oct.} \bibinfo{year}{2020}).
\newblock


\bibitem[\protect\citeauthoryear{Grimmelmann}{Grimmelmann}{2015}]%
        {grimmelmann_virtues_2015}
\bibfield{author}{\bibinfo{person}{James Grimmelmann}.}
  \bibinfo{year}{2015}\natexlab{}.
\newblock \showarticletitle{The {{Virtues}} of {{Moderation}}}.
\newblock \bibinfo{journal}{\emph{Yale Journal of Law and Technology}}
  \bibinfo{volume}{17}, \bibinfo{number}{1} (\bibinfo{date}{Sept.}
  \bibinfo{year}{2015}).
\newblock


\bibitem[\protect\citeauthoryear{Gurzick, White, Lutters, and Boot}{Gurzick
  et~al\mbox{.}}{2009}]%
        {gurzick_view_2009}
\bibfield{author}{\bibinfo{person}{David Gurzick}, \bibinfo{person}{Kevin~F.
  White}, \bibinfo{person}{Wayne~G. Lutters}, {and} \bibinfo{person}{Lee
  Boot}.} \bibinfo{year}{2009}\natexlab{}.
\newblock \showarticletitle{A View from {{Mount Olympus}}: The Impact of
  Activity Tracking Tools on the Character and Practice of Moderation}. In
  \bibinfo{booktitle}{\emph{Proceedings of {{GROUP}}}}.
\newblock


\bibitem[\protect\citeauthoryear{Halfaker and Geiger}{Halfaker and
  Geiger}{2020}]%
        {halfaker_ores_2020}
\bibfield{author}{\bibinfo{person}{Aaron Halfaker} {and}
  \bibinfo{person}{R.~Stuart Geiger}.} \bibinfo{year}{2020}\natexlab{}.
\newblock \showarticletitle{{{ORES}}: {{Lowering Barriers}} with
  {{Participatory Machine Learning}} in {{Wikipedia}}}. In
  \bibinfo{booktitle}{\emph{Proceedings of {{CSCW}}}}.
\newblock


\bibitem[\protect\citeauthoryear{Halfaker, Geiger, Morgan, and Riedl}{Halfaker
  et~al\mbox{.}}{2013}]%
        {halfaker_rise_2013}
\bibfield{author}{\bibinfo{person}{Aaron Halfaker}, \bibinfo{person}{R.~Stuart
  Geiger}, \bibinfo{person}{Jonathan~T. Morgan}, {and} \bibinfo{person}{John
  Riedl}.} \bibinfo{year}{2013}\natexlab{}.
\newblock \showarticletitle{The {{Rise}} and {{Decline}} of an {{Open
  Collaboration System}}}.
\newblock \bibinfo{journal}{\emph{American Behavioral Scientist}}
  \bibinfo{volume}{57}, \bibinfo{number}{5} (\bibinfo{date}{May}
  \bibinfo{year}{2013}).
\newblock


\bibitem[\protect\citeauthoryear{Halfaker, Kittur, and Riedl}{Halfaker
  et~al\mbox{.}}{2011a}]%
        {halfaker_dont_2011}
\bibfield{author}{\bibinfo{person}{Aaron Halfaker}, \bibinfo{person}{Aniket
  Kittur}, {and} \bibinfo{person}{John Riedl}.}
  \bibinfo{year}{2011}\natexlab{a}.
\newblock \showarticletitle{Don't Bite the Newbies: How Reverts Affect the
  Quantity and Quality of {{Wikipedia}} Work}. In
  \bibinfo{booktitle}{\emph{Proceedings of {{WikiSym}}}}.
\newblock


\bibitem[\protect\citeauthoryear{Halfaker, Song, Stuart, Kittur, and
  Riedl}{Halfaker et~al\mbox{.}}{2011b}]%
        {halfaker_nice:_2011}
\bibfield{author}{\bibinfo{person}{Aaron Halfaker}, \bibinfo{person}{Bryan
  Song}, \bibinfo{person}{D.~Alex Stuart}, \bibinfo{person}{Aniket Kittur},
  {and} \bibinfo{person}{John Riedl}.} \bibinfo{year}{2011}\natexlab{b}.
\newblock \showarticletitle{{{NICE}}: {{Social Translucence Through UI
  Intervention}}}. In \bibinfo{booktitle}{\emph{Proceedings of {{WikiSym}}}}.
\newblock


\bibitem[\protect\citeauthoryear{Henner and Sefidari}{Henner and
  Sefidari}{2016}]%
        {henner_wikimedia_2016}
\bibfield{author}{\bibinfo{person}{Christophe Henner} {and}
  \bibinfo{person}{Maria Sefidari}.} \bibinfo{year}{2016}\natexlab{}.
\newblock \bibinfo{title}{Wikimedia {{Foundation Board}} on Healthy
  {{Wikimedia}} Community Culture, Inclusivity, and Safe Spaces \textendash{}
  {{Wikimedia Blog}}}.
\newblock
\newblock
\urldef\tempurl%
\url{https://blog.wikimedia.org/2016/12/08/board-culture-inclusivity-safe-spaces/}
\showURL{%
\tempurl}


\bibitem[\protect\citeauthoryear{Jagannath, Salen, and Slov{\`a}k}{Jagannath
  et~al\mbox{.}}{2020}]%
        {jagannath_we_2020}
\bibfield{author}{\bibinfo{person}{Krithika Jagannath}, \bibinfo{person}{Katie
  Salen}, {and} \bibinfo{person}{Petr Slov{\`a}k}.}
  \bibinfo{year}{2020}\natexlab{}.
\newblock \showarticletitle{"({{We}}) {{Can Talk It Out}}...": {{Designing}}
  for {{Promoting Conflict-Resolution Skills}} in {{Youth}} on a {{Moderated
  Minecraft Server}}}. In \bibinfo{booktitle}{\emph{Proceedings of {{CSCW}}}}.
\newblock


\bibitem[\protect\citeauthoryear{Jhaver, Appling, Gilbert, and Bruckman}{Jhaver
  et~al\mbox{.}}{2019a}]%
        {jhaver_did_2019}
\bibfield{author}{\bibinfo{person}{Shagun Jhaver},
  \bibinfo{person}{Darren~Scott Appling}, \bibinfo{person}{Eric Gilbert}, {and}
  \bibinfo{person}{Amy Bruckman}.} \bibinfo{year}{2019}\natexlab{a}.
\newblock \showarticletitle{``{{Did You Suspect}} the {{Post Would}} Be
  {{Removed}}?'': {{Understanding User Reactions}} to {{Content Removals}} on
  {{Reddit}}}. In \bibinfo{booktitle}{\emph{Proceedings of {{CSCW}}}}.
\newblock


\bibitem[\protect\citeauthoryear{Jhaver, Birman, Gilbert, and Bruckman}{Jhaver
  et~al\mbox{.}}{2019b}]%
        {jhaver_human-machine_2019}
\bibfield{author}{\bibinfo{person}{Shagun Jhaver}, \bibinfo{person}{Iris
  Birman}, \bibinfo{person}{Eric Gilbert}, {and} \bibinfo{person}{Amy
  Bruckman}.} \bibinfo{year}{2019}\natexlab{b}.
\newblock \showarticletitle{Human-{{Machine Collaboration}} for {{Content
  Regulation}}: {{The Case}} of {{Reddit Automoderator}}}.
\newblock \bibinfo{journal}{\emph{ACM Transactions on Computer-Human
  Interaction}} \bibinfo{volume}{26}, \bibinfo{number}{5} (\bibinfo{date}{July}
  \bibinfo{year}{2019}).
\newblock


\bibitem[\protect\citeauthoryear{Jhaver, Boylston, Yang, and Bruckman}{Jhaver
  et~al\mbox{.}}{2021}]%
        {jhaver_evaluating_2021}
\bibfield{author}{\bibinfo{person}{Shagun Jhaver}, \bibinfo{person}{Christian
  Boylston}, \bibinfo{person}{Diyi Yang}, {and} \bibinfo{person}{Amy
  Bruckman}.} \bibinfo{year}{2021}\natexlab{}.
\newblock \showarticletitle{Evaluating the {{Effectiveness}} of
  {{Deplatforming}} as a {{Moderation Strategy}} on {{Twitter}}}. In
  \bibinfo{booktitle}{\emph{Proceedings of {{CSCW}}}}.
\newblock


\bibitem[\protect\citeauthoryear{Jhaver, Ghoshal, Bruckman, and Gilbert}{Jhaver
  et~al\mbox{.}}{2018}]%
        {jhaver_online_2018}
\bibfield{author}{\bibinfo{person}{Shagun Jhaver}, \bibinfo{person}{Sucheta
  Ghoshal}, \bibinfo{person}{Amy Bruckman}, {and} \bibinfo{person}{Eric
  Gilbert}.} \bibinfo{year}{2018}\natexlab{}.
\newblock \showarticletitle{Online {{Harassment}} and {{Content Moderation}}:
  {{The Case}} of {{Blocklists}}}.
\newblock \bibinfo{journal}{\emph{ACM Transactions on Computer-Human
  Interaction}} \bibinfo{volume}{25}, \bibinfo{number}{2}
  (\bibinfo{date}{March} \bibinfo{year}{2018}).
\newblock


\bibitem[\protect\citeauthoryear{Johnson}{Johnson}{2017}]%
        {johnson_multiple_2017}
\bibfield{author}{\bibinfo{person}{Amy Johnson}.}
  \bibinfo{year}{2017}\natexlab{}.
\newblock \showarticletitle{The {{Multiple Harms}} of {{Sea Lions}}}.
\newblock In \bibinfo{booktitle}{\emph{Perspectives on {{Harmful Speech
  Online}}.}} \bibinfo{publisher}{{Berkman Klein Center for Internet \&
  Society.}}
\newblock


\bibitem[\protect\citeauthoryear{Jurgens, Hemphill, and
  Chandrasekharan}{Jurgens et~al\mbox{.}}{2019}]%
        {jurgens_just_2019}
\bibfield{author}{\bibinfo{person}{David Jurgens}, \bibinfo{person}{Libby
  Hemphill}, {and} \bibinfo{person}{Eshwar Chandrasekharan}.}
  \bibinfo{year}{2019}\natexlab{}.
\newblock \showarticletitle{A {{Just}} and {{Comprehensive Strategy}} for
  {{Using NLP}} to {{Address Online Abuse}}}. In
  \bibinfo{booktitle}{\emph{Proceedings of {{ACL}}}}.
\newblock


\bibitem[\protect\citeauthoryear{Kiesler, Kraut, Resnick, and Kittur}{Kiesler
  et~al\mbox{.}}{2012}]%
        {kiesler_regulating_2012}
\bibfield{author}{\bibinfo{person}{Sara Kiesler}, \bibinfo{person}{Robert
  Kraut}, \bibinfo{person}{Paul Resnick}, {and} \bibinfo{person}{Aniket
  Kittur}.} \bibinfo{year}{2012}\natexlab{}.
\newblock \showarticletitle{Regulating Behavior in Online Communities.}
\newblock In \bibinfo{booktitle}{\emph{Building {{Successful Online
  Communities}}: {{Evidence-Based Social Design}}}},
  \bibfield{editor}{\bibinfo{person}{Paul Resnick} {and}
  \bibinfo{person}{Robert Kraut}} (Eds.). \bibinfo{publisher}{{MIT Press}}.
\newblock


\bibitem[\protect\citeauthoryear{Kittur and Kraut}{Kittur and Kraut}{2008}]%
        {kittur_harnessing_2008}
\bibfield{author}{\bibinfo{person}{Aniket Kittur} {and}
  \bibinfo{person}{Robert~E. Kraut}.} \bibinfo{year}{2008}\natexlab{}.
\newblock \showarticletitle{Harnessing the {{Wisdom}} of {{Crowds}} in
  {{Wikipedia}}: {{Quality Through Coordination}}}. In
  \bibinfo{booktitle}{\emph{Proceedings of {{CSCW}}}}.
\newblock


\bibitem[\protect\citeauthoryear{Kittur, Suh, Pendleton, and Chi}{Kittur
  et~al\mbox{.}}{2007}]%
        {kittur_he_2007}
\bibfield{author}{\bibinfo{person}{Aniket Kittur}, \bibinfo{person}{Bongwon
  Suh}, \bibinfo{person}{Bryan~A. Pendleton}, {and} \bibinfo{person}{Ed~H.
  Chi}.} \bibinfo{year}{2007}\natexlab{}.
\newblock \showarticletitle{He Says, She Says: Conflict and Coordination in
  {{Wikipedia}}}. In \bibinfo{booktitle}{\emph{Proceedings of {{CHI}}}}.
\newblock


\bibitem[\protect\citeauthoryear{Kriplean, Morgan, Freelon, Borning, and
  Bennett}{Kriplean et~al\mbox{.}}{2012a}]%
        {kriplean_supporting_2012}
\bibfield{author}{\bibinfo{person}{Travis Kriplean}, \bibinfo{person}{Jonathan
  Morgan}, \bibinfo{person}{Deen Freelon}, \bibinfo{person}{Alan Borning},
  {and} \bibinfo{person}{Lance Bennett}.} \bibinfo{year}{2012}\natexlab{a}.
\newblock \showarticletitle{Supporting Reflective Public Thought with
  Considerit}. In \bibinfo{booktitle}{\emph{Proceedings of {{CSCW}}}}.
\newblock


\bibitem[\protect\citeauthoryear{Kriplean, Toomim, Morgan, Borning, and
  Ko}{Kriplean et~al\mbox{.}}{2012b}]%
        {kriplean_is_2012}
\bibfield{author}{\bibinfo{person}{Travis Kriplean}, \bibinfo{person}{Michael
  Toomim}, \bibinfo{person}{Jonathan Morgan}, \bibinfo{person}{Alan Borning},
  {and} \bibinfo{person}{Andrew Ko}.} \bibinfo{year}{2012}\natexlab{b}.
\newblock \showarticletitle{Is This What You Meant?: Promoting Listening on the
  Web with Reflect}. In \bibinfo{booktitle}{\emph{Proceedings of {{CHI}}}}.
\newblock


\bibitem[\protect\citeauthoryear{Lampe and Resnick}{Lampe and Resnick}{2004}]%
        {lampe_slashdot_2004}
\bibfield{author}{\bibinfo{person}{Cliff Lampe} {and} \bibinfo{person}{Paul
  Resnick}.} \bibinfo{year}{2004}\natexlab{}.
\newblock \showarticletitle{Slash(Dot) and Burn: Distributed Moderation in a
  Large Online Conversation Space}. In \bibinfo{booktitle}{\emph{Proceedings of
  {{CHI}}}}.
\newblock


\bibitem[\protect\citeauthoryear{Liu, Guberman, Hemphill, and Culotta}{Liu
  et~al\mbox{.}}{2018}]%
        {liu_forecasting_2018}
\bibfield{author}{\bibinfo{person}{Ping Liu}, \bibinfo{person}{Joshua
  Guberman}, \bibinfo{person}{Libby Hemphill}, {and} \bibinfo{person}{Aron
  Culotta}.} \bibinfo{year}{2018}\natexlab{}.
\newblock \showarticletitle{Forecasting the {{Presence}} and {{Intensity}} of
  {{Hostility}} on {{Instagram Using Linguistic}} and {{Social Features}}}. In
  \bibinfo{booktitle}{\emph{Proceedings of {{ICWSM}}}}.
\newblock


\bibitem[\protect\citeauthoryear{Lo}{Lo}{2018}]%
        {lo_when_2018}
\bibfield{author}{\bibinfo{person}{Claudia (Claudia Wai~Yu) Lo}.}
  \bibinfo{year}{2018}\natexlab{}.
\newblock \emph{\bibinfo{title}{When All You Have Is a Banhammer : The Social
  and Communicative Work of {{Volunteer}} Moderators}}.
\newblock Thesis. \bibinfo{school}{Massachusetts Institute of Technology}.
\newblock


\bibitem[\protect\citeauthoryear{Matias}{Matias}{2019}]%
        {matias_civic_2019}
\bibfield{author}{\bibinfo{person}{J.~Nathan Matias}.}
  \bibinfo{year}{2019}\natexlab{}.
\newblock \showarticletitle{The {{Civic Labor}} of {{Volunteer Moderators
  Online}}}.
\newblock \bibinfo{journal}{\emph{Social Media + Society}} \bibinfo{volume}{5},
  \bibinfo{number}{2} (\bibinfo{date}{April} \bibinfo{year}{2019}).
\newblock


\bibitem[\protect\citeauthoryear{McGillicuddy, Bernard, and
  Cranefield}{McGillicuddy et~al\mbox{.}}{2016}]%
        {mcgillicuddy_controlling_2016}
\bibfield{author}{\bibinfo{person}{Aiden~R McGillicuddy},
  \bibinfo{person}{Jean-Gregoire Bernard}, {and} \bibinfo{person}{Jocelyn~Ann
  Cranefield}.} \bibinfo{year}{2016}\natexlab{}.
\newblock \showarticletitle{Controlling {{Bad Behavior}} in {{Online
  Communities}}: {{An Examination}} of {{Moderation Work}}}. In
  \bibinfo{booktitle}{\emph{Proceedings of {{ICIS}}}}.
\newblock


\bibitem[\protect\citeauthoryear{Morgan and Halfaker}{Morgan and
  Halfaker}{2018}]%
        {morgan_evaluating_2018}
\bibfield{author}{\bibinfo{person}{Jonathan~T. Morgan} {and}
  \bibinfo{person}{Aaron Halfaker}.} \bibinfo{year}{2018}\natexlab{}.
\newblock \showarticletitle{Evaluating the Impact of the {{Wikipedia Teahouse}}
  on Newcomer Socialization and Retention}. In
  \bibinfo{booktitle}{\emph{Proceedings of {{OpenSym}}}}.
\newblock


\bibitem[\protect\citeauthoryear{Nobata, Tetreault, Thomas, Mehdad, and
  Chang}{Nobata et~al\mbox{.}}{2016}]%
        {nobata_abusive_2016}
\bibfield{author}{\bibinfo{person}{Chikashi Nobata}, \bibinfo{person}{Joel
  Tetreault}, \bibinfo{person}{Achint Thomas}, \bibinfo{person}{Yashar Mehdad},
  {and} \bibinfo{person}{Yi Chang}.} \bibinfo{year}{2016}\natexlab{}.
\newblock \showarticletitle{Abusive {{Language Detection}} in {{Online User
  Content}}}. In \bibinfo{booktitle}{\emph{Proceedings of {{WWW}}}}.
\newblock


\bibitem[\protect\citeauthoryear{Park, Sachar, Diakopoulos, and Elmqvist}{Park
  et~al\mbox{.}}{2016}]%
        {park_supporting_2016}
\bibfield{author}{\bibinfo{person}{Deokgun Park}, \bibinfo{person}{Simranjit
  Sachar}, \bibinfo{person}{Nicholas Diakopoulos}, {and}
  \bibinfo{person}{Niklas Elmqvist}.} \bibinfo{year}{2016}\natexlab{}.
\newblock \showarticletitle{Supporting {{Comment Moderators}} in {{Identifying
  High Quality Online News Comments}}}. In
  \bibinfo{booktitle}{\emph{Proceedings of {{CHI}}}}.
\newblock


\bibitem[\protect\citeauthoryear{Raskauskas and Stoltz}{Raskauskas and
  Stoltz}{2007}]%
        {raskauskas_involvement_2007}
\bibfield{author}{\bibinfo{person}{Juliana Raskauskas} {and}
  \bibinfo{person}{Ann~D. Stoltz}.} \bibinfo{year}{2007}\natexlab{}.
\newblock \showarticletitle{Involvement in {{Traditional}} and {{Electronic
  Bullying Among Adolescents}}}.
\newblock \bibinfo{journal}{\emph{Developmental Psychology}}
  \bibinfo{volume}{43}, \bibinfo{number}{3} (\bibinfo{date}{May}
  \bibinfo{year}{2007}).
\newblock


\bibitem[\protect\citeauthoryear{Sap, Card, Gabriel, Choi, and Smith}{Sap
  et~al\mbox{.}}{2019}]%
        {sap_risk_2019}
\bibfield{author}{\bibinfo{person}{Maarten Sap}, \bibinfo{person}{Dallas Card},
  \bibinfo{person}{Saadia Gabriel}, \bibinfo{person}{Yejin Choi}, {and}
  \bibinfo{person}{Noah~A. Smith}.} \bibinfo{year}{2019}\natexlab{}.
\newblock \showarticletitle{The {{Risk}} of {{Racial Bias}} in {{Hate Speech
  Detection}}}. In \bibinfo{booktitle}{\emph{Proceedings of {{ACL}}}}.
\newblock


\bibitem[\protect\citeauthoryear{Seering}{Seering}{2020}]%
        {seering_reconsidering_2020}
\bibfield{author}{\bibinfo{person}{Joseph Seering}.}
  \bibinfo{year}{2020}\natexlab{}.
\newblock \showarticletitle{Reconsidering {{Self-Moderation}}: The {{Role}} of
  {{Research}} in {{Supporting Community-Based Models}} for {{Online Content
  Moderation}}}. In \bibinfo{booktitle}{\emph{Proceedings of {{CSCW}}}}.
\newblock


\bibitem[\protect\citeauthoryear{Seering, Fang, Damasco, Chen, Sun, and
  Kaufman}{Seering et~al\mbox{.}}{2019a}]%
        {seering_designing_2019}
\bibfield{author}{\bibinfo{person}{Joseph Seering}, \bibinfo{person}{Tianmi
  Fang}, \bibinfo{person}{Luca Damasco}, \bibinfo{person}{Mianhong~'Cherie'
  Chen}, \bibinfo{person}{Likang Sun}, {and} \bibinfo{person}{Geoff Kaufman}.}
  \bibinfo{year}{2019}\natexlab{a}.
\newblock \showarticletitle{Designing {{User Interface Elements}} to
  {{Improve}} the {{Quality}} and {{Civility}} of {{Discourse}} in {{Online
  Commenting Behaviors}}}. In \bibinfo{booktitle}{\emph{Proceedings of
  {{CHI}}}}.
\newblock


\bibitem[\protect\citeauthoryear{Seering, Kaufman, and Chancellor}{Seering
  et~al\mbox{.}}{2020}]%
        {seering_metaphors_2020}
\bibfield{author}{\bibinfo{person}{Joseph Seering}, \bibinfo{person}{Geoff
  Kaufman}, {and} \bibinfo{person}{Stevie Chancellor}.}
  \bibinfo{year}{2020}\natexlab{}.
\newblock \showarticletitle{Metaphors in Moderation}.
\newblock \bibinfo{journal}{\emph{New Media \& Society}} (\bibinfo{date}{Oct.}
  \bibinfo{year}{2020}).
\newblock


\bibitem[\protect\citeauthoryear{Seering, Kraut, and Dabbish}{Seering
  et~al\mbox{.}}{2017}]%
        {seering_shaping_2017}
\bibfield{author}{\bibinfo{person}{Joseph Seering}, \bibinfo{person}{Robert
  Kraut}, {and} \bibinfo{person}{Laura Dabbish}.}
  \bibinfo{year}{2017}\natexlab{}.
\newblock \showarticletitle{Shaping {{Pro}} and {{Anti-Social Behavior}} on
  {{Twitch Through Moderation}} and {{Example-Setting}}}. In
  \bibinfo{booktitle}{\emph{Proceedings of {{CSCW}}}}.
\newblock


\bibitem[\protect\citeauthoryear{Seering, Wang, Yoon, and Kaufman}{Seering
  et~al\mbox{.}}{2019b}]%
        {seering_moderator_2019}
\bibfield{author}{\bibinfo{person}{Joseph Seering}, \bibinfo{person}{Tony
  Wang}, \bibinfo{person}{Jina Yoon}, {and} \bibinfo{person}{Geoff Kaufman}.}
  \bibinfo{year}{2019}\natexlab{b}.
\newblock \showarticletitle{Moderator Engagement and Community Development in
  the Age of Algorithms}.
\newblock \bibinfo{journal}{\emph{New Media \& Society}} \bibinfo{volume}{21},
  \bibinfo{number}{7} (\bibinfo{date}{July} \bibinfo{year}{2019}).
\newblock


\bibitem[\protect\citeauthoryear{Srinivasan, {Danescu-Niculescu-Mizil}, Lee,
  and Tan}{Srinivasan et~al\mbox{.}}{2019}]%
        {srinivasan_content_2019}
\bibfield{author}{\bibinfo{person}{Kumar Srinivasan}, \bibinfo{person}{Cristian
  {Danescu-Niculescu-Mizil}}, \bibinfo{person}{Lillian Lee}, {and}
  \bibinfo{person}{Chenhao Tan}.} \bibinfo{year}{2019}\natexlab{}.
\newblock \showarticletitle{Content {{Removal}} as a {{Moderation Strategy}}:
  {{Compliance}} and {{Other Outcomes}} in the {{ChangeMyView Community}}}. In
  \bibinfo{booktitle}{\emph{Proceedings of {{CSCW}}}}.
\newblock


\bibitem[\protect\citeauthoryear{Taylor, DiFranzo, Choi, Sannon, and
  Bazarova}{Taylor et~al\mbox{.}}{2019}]%
        {taylor_accountability_2019}
\bibfield{author}{\bibinfo{person}{Samuel~Hardman Taylor},
  \bibinfo{person}{Dominic DiFranzo}, \bibinfo{person}{Yoon~Hyung Choi},
  \bibinfo{person}{Shruti Sannon}, {and} \bibinfo{person}{Natalya~N.
  Bazarova}.} \bibinfo{year}{2019}\natexlab{}.
\newblock \showarticletitle{Accountability and {{Empathy}} by {{Design}}:
  {{Encouraging Bystander Intervention}} to {{Cyberbullying}} on {{Social
  Media}}}. In \bibinfo{booktitle}{\emph{Proceedings of {{CSCW}}}}.
\newblock


\bibitem[\protect\citeauthoryear{Wiegand, Ruppenhofer, and Kleinbauer}{Wiegand
  et~al\mbox{.}}{2019}]%
        {wiegand_detection_2019}
\bibfield{author}{\bibinfo{person}{Michael Wiegand}, \bibinfo{person}{Josef
  Ruppenhofer}, {and} \bibinfo{person}{Thomas Kleinbauer}.}
  \bibinfo{year}{2019}\natexlab{}.
\newblock \showarticletitle{Detection of {{Abusive Language}}: The {{Problem}}
  of {{Biased Datasets}}}. In \bibinfo{booktitle}{\emph{Proceedings of
  {{NAACL}}}}.
\newblock


\bibitem[\protect\citeauthoryear{{Wikimedia Support {and} Safety
  Team}}{{Wikimedia Support {and} Safety Team}}{2015}]%
        {wikimedia_support_and_safety_team_harassment_2015}
\bibfield{author}{\bibinfo{person}{{Wikimedia Support {and} Safety Team}}.}
  \bibinfo{year}{2015}\natexlab{}.
\newblock \bibinfo{title}{Harassment {{Survey}}}.
\newblock
\newblock
\urldef\tempurl%
\url{https://upload.wikimedia.org/wikipedia/commons/5/52/Harassment\_Survey\_2015\_-\_Results\_Report.pdf}
\showURL{%
\tempurl}


\bibitem[\protect\citeauthoryear{Wohn}{Wohn}{2019}]%
        {wohn_volunteer_2019}
\bibfield{author}{\bibinfo{person}{Donghee~Yvette Wohn}.}
  \bibinfo{year}{2019}\natexlab{}.
\newblock \showarticletitle{Volunteer {{Moderators}} in {{Twitch Micro
  Communities}}: {{How They Get Involved}}, the {{Roles They Play}}, and the
  {{Emotional Labor They Experience}}}. In
  \bibinfo{booktitle}{\emph{Proceedings of {{CHI}}}}.
\newblock


\bibitem[\protect\citeauthoryear{Wulczyn, Thain, and Dixon}{Wulczyn
  et~al\mbox{.}}{2017}]%
        {wulczyn_ex_2017}
\bibfield{author}{\bibinfo{person}{Ellery Wulczyn}, \bibinfo{person}{Nithum
  Thain}, {and} \bibinfo{person}{Lucas Dixon}.}
  \bibinfo{year}{2017}\natexlab{}.
\newblock \showarticletitle{Ex {{Machina}}: {{Personal Attacks Seen}} at
  {{Scale}}}. In \bibinfo{booktitle}{\emph{Proceedings of {{WWW}}}}.
\newblock


\bibitem[\protect\citeauthoryear{Zhang, Chang, {Danescu-Niculescu-Mizil},
  Dixon, Thain, Hua, and Taraborelli}{Zhang et~al\mbox{.}}{2018}]%
        {zhang_conversations_2018}
\bibfield{author}{\bibinfo{person}{Justine Zhang}, \bibinfo{person}{Jonathan~P.
  Chang}, \bibinfo{person}{Cristian {Danescu-Niculescu-Mizil}},
  \bibinfo{person}{Lucas Dixon}, \bibinfo{person}{Nithum Thain},
  \bibinfo{person}{Yiqing Hua}, {and} \bibinfo{person}{Dario Taraborelli}.}
  \bibinfo{year}{2018}\natexlab{}.
\newblock \showarticletitle{Conversations {{Gone Awry}}: {{Detecting Early
  Signs}} of {{Conversational Failure}}}. In
  \bibinfo{booktitle}{\emph{Proceedings of {{ACL}}}}.
\newblock


\bibitem[\protect\citeauthoryear{Zheng, Albano, Vora, Mai, and Nickerson}{Zheng
  et~al\mbox{.}}{2019}]%
        {zheng_roles_2019}
\bibfield{author}{\bibinfo{person}{Lei~(Nico) Zheng},
  \bibinfo{person}{Christopher~M. Albano}, \bibinfo{person}{Neev~M. Vora},
  \bibinfo{person}{Feng Mai}, {and} \bibinfo{person}{Jeffrey~V. Nickerson}.}
  \bibinfo{year}{2019}\natexlab{}.
\newblock \showarticletitle{The {{Roles Bots Play}} in {{Wikipedia}}}. In
  \bibinfo{booktitle}{\emph{Proceedings of {{CSCW}}}}.
\newblock


\end{thebibliography}

\appendix

\section{Interview Questions}
\label{apdx:interview_qs}

This appendix shows the general outline we followed for all moderator interviews. Note that this only served as a general guide; as the interview process is semi-structured we let the conversation flow naturally, so the exact order and wording of questions varied in practice.

\subsection*{Topic 1: Current Discussion Moderation Practices}
\begin{itemize}
    \item Understanding comment removal practices:
    \begin{itemize}
        \item Q: How do you select comments to inspect for incivility and community rule violations?
        \item Q: Do you ever proactively monitor ongoing conversations that you consider to be at risk of derailing into uncivil behavior?
        \item Q: Say that you have a potentially problematic comment. Please describe your typical process for determining whether or not this comment needs moderation action.
        \begin{itemize}
            \item (Optional) Q: Can you think of a specific example of a comment you took action on, and describe the process of determining whether or not that comment needed moderation action?
        \end{itemize}
    \end{itemize}
    \item Understanding how moderators use context:
    \begin{itemize}
        \item Q: When you are considering [moderating/removing] a comment, do you generally read earlier comments in the thread for context?  If so, what are you looking for?
        \item Q: Do you think a user’s [post and comment/edit] history affects your decision on whether you remove one of their comments?  Can you give examples of such historical factors?
        \item Q: After you take some moderation action on a comment, would you ever look at earlier comments in the conversation to identify more rule-violating comments?
    \end{itemize}
    \item Understanding how moderators use automated tools:
    \begin{itemize}
        \item Q: What automated tools do you currently use for moderation, if any? 
        \item Q: If you use any automated tools, do you use them for
        \begin{itemize}
            \item triaging comments for you to review
            \item \textit{and/or} 
            \item automatically removing content?
            \item If so, how do you configure automod for your community?
        \end{itemize}
    \end{itemize}
    \item Moderators’ motivation and how they view the role of moderator:
    \begin{itemize}
        \item Q: Why did you become a [moderator/administrator] for [Reddit/Wikipedia]?
        \item Q: As an administrator, why did you become involved in discussion moderation work?
    \end{itemize}
    \item Miscellaneous:
    \begin{itemize}
        \item Q: How much time do you spend moderating each day? Each week?
        \begin{itemize}
            \item First ask about time spent as an administrator, then ask about moderation. 
        \end{itemize}
        \item Q: How satisfied are you with your current moderation practices? Do you see room for improvement? 
        \item Q: When doing your job as a moderator, would you rather:
        \begin{itemize}
            \item (a) only [remove/take moderation action] very flagrant rule violations, and potentially miss some rule-breaking comments, or
            \item (b) [remove/take moderation action] all comments that could be rule violations, potentially  [remove/take moderation action] some comments that don’t deserve to be. 
        \end{itemize}
    \end{itemize}
\end{itemize}

\subsection*{Topic 2: Potential Use of Conversational Forecasting}
\begin{itemize}
    \item Understanding what moderators would do without time constraints:
    \begin{itemize}
        \item Q: If you had more time for your job as a moderator, what actions would you want to do?
        \item Q: If you had more time for your job as a moderator, when you make a moderation decision about a comment, would you read more of the context around the comment to inform your decision?
    \end{itemize}
    \item Can moderators tell if a conversation is going awry? Can anyone?
    \begin{itemize}
        \item Explanation: Here is some terminology that we will use for the rest of the interview:
        \begin{itemize}
            \item We'll say that a comment is \textit{civil} if it follows all the rules of your community, and that it is \textit{uncivil} if it violates a community rule. 
            \item We will also say that a conversation \textit{eventually derails} if it is civil right now, but in the future an uncivil comment gets posted to the conversation.
        \end{itemize}
        \item Q: Given a civil conversation, do you think it is possible to foretell if a conversation will \textit{eventually derail} into uncivil comments?
        \item Q: Do you think you yourself are able to do this prediction?
        \begin{itemize}
            \item If yes:
            \begin{itemize}
                \item Q: Roughly how often do you think your prediction would be correct? That is, can you estimate what portion of the conversations you think will derail actually do end up derailing?
                \item Q: What clues from a conversation do you use to inform your prediction? 
            \end{itemize}
        \end{itemize}
        \item Q: Do you think other moderators would be able to do this type of prediction? 
        \item Q: Do you think an algorithm might be able to do this type of prediction?
        \begin{itemize}
            \item Q: Do you think it would be better or worse than humans?
        \end{itemize}
    \end{itemize}
    \item Monitoring derailing conversations:
    \begin{itemize}
        \item Q: Assume you would know \textit{for sure} that an ongoing conversation will turn uncivil in the future. Would you like to monitor new comments that are posted in this conversation? 
        \item Q: Now consider a more realistic scenario, where you cannot know for sure what the future of a conversation will be. Now, say we have a conversation that is predicted to derail; we will go through various levels of confidence in this prediction, and I want you to tell me if you would want to monitor new comments in the conversation for each level of prediction confidence.
        \begin{itemize}
            \item Would you want to monitor new comments if you had \textit{low} certainty in the prediction (i.e., \textit{20\%} of the conversations that are predicted to derail will eventually actually end up derailing)?
            \item Would you want to monitor new comments if you had \textit{50-50} certainty in the prediction  (i.e., \textit{50\%} of the conversations that are predicted to derail will eventually actually end up derailing)?
            \item Would you want to monitor new comments if you had \textit{high} certainty in the prediction (i.e., \textit{80\%} of the conversations that are predicted to derail will eventually actually end up derailing)?
        \end{itemize}
    \end{itemize}
    \item Taking proactive steps for derailing conversations
    \begin{itemize}
        \item Q: Assuming you would know for sure that a (currently civil) conversation will turn uncivil and violate the rules of the community, what proactive steps do you, as a moderator, see yourself taking in order to prevent uncivil behavior (if any)?
        \item Q: Now---as before---consider a more realistic scenario, where you cannot know for sure what the future of a conversation will be. Now, say we have a conversation that is predicted to derail; we will go through various levels of confidence in this prediction, and I want you to tell me which of the proactive steps you just mentioned you would still take for each level of prediction confidence.
        \begin{itemize}
            \item What proactive steps would you take if you had \textit{low} certainty in the prediction (i.e., \textit{20\%} of the conversations that are predicted to derail will eventually actually end up derailing)?
            \item What proactive steps would you take if you had \textit{50-50} certainty in the prediction  (i.e., \textit{50\%} of the conversations that are predicted to derail will eventually actually end up derailing)?
            \item What proactive steps would you take if you had \textit{high} certainty in the prediction (i.e., \textit{80\%} of the conversations that are predicted to derail will eventually actually end up derailing)?
        \end{itemize}
        \item Q: Have you ever taken any of these proactive steps in the past?
    \end{itemize}
\end{itemize}

\subsection*{Topic 3: Analyzing a Mockup Conversation}
[The participant is shown a conversation from their community in the Conversation View, with the CRAFT score annotations removed.]
\begin{itemize}
    \item Q: Do you think any comments in this conversation are uncivil?
    \item Q: How likely do you think this conversation is to eventually derail into uncivil behavior (breaking the rules of the community)?  What made you think this way (point to specific behaviors)?
    \item Q: Would you want to monitor new comments in this conversation?    
    \item Q: Would you consider taking any proactive steps to prevent uncivil behavior?
\end{itemize}

\noindent[The participant is shown the CRAFT scores for this conversation.]

\begin{itemize}
    \item Ask the same questions again.
\end{itemize}

\subsection*{Topic 4: Analyzing a Mockup Ranking}
[The moderator is shown a ranking of conversations from their community in the Ranking View.]
\begin{itemize}
    \item Q: Which conversations do you think would be worth monitoring for uncivil behavior?
    \item Q: On the main page for the listing, how relevant is the information displayed about each thread? 
    \item Q: What other information would you find useful in deciding whether to inspect or monitor a conversation?
\end{itemize}

\end{document}